\documentclass[aps,twocolumn,superscriptaddress,showpacs]{revtex4}
\usepackage{amssymb,amsmath,epsfig,lscape,color,eucal,graphicx,bm,epsfig,multirow}

\def\av#1{\left\langle#1\right\rangle}

\usepackage{latexsym}
\usepackage{amstext,amssymb,amsfonts}
\usepackage{graphicx}
\usepackage{color}
\usepackage{subfigure}

\begin{document}

\title{Random walk with priorities in communication-like networks}

\author{Nikolaos Bastas}
\affiliation{Department of Physics, University of Thessaloniki,54124 Thessaloniki, Greece}
\author{Michalis Maragakis}
\affiliation{Department of Physics, University of Thessaloniki,54124 Thessaloniki, Greece}
\affiliation{Department of Economics, University of Macedonia, 54006 Thessaloniki, Greece}
\author{Panos Argyrakis}
\affiliation{Department of Physics, University of Thessaloniki,54124 Thessaloniki, Greece}
\author{Daniel ben-Avraham}
\affiliation{Department of Physics, Clarkson University, Potsdam, New York 13699-5820, USA}
\author{Shlomo Havlin}
\affiliation{Department of Physics, Bar-Ilan University, Ramat Gan 52900, Israel}
\author{Shai Carmi}
\affiliation{Department of Physics, Bar-Ilan University, Ramat Gan 52900, Israel}

\date{\today}

\begin{abstract}
We study a model for a random walk of two classes of particles ($A$ and $B$). Where both species are present in the same site, the motion of $A$'s takes precedence over that of $B$'s. The model was originally proposed and analyzed in Maragakis et al., \emph{Phys. Rev. E} \textbf{77}, 020103 (2008); here we provide additional results. We solve analytically the diffusion coefficients of the two species in lattices for a number of protocols. In networks, we find that the probability of a $B$ particle to be free decreases exponentially with the node degree. In scale-free networks, this leads to localization of the $B$'s at the hubs and arrest of their motion. To remedy this, we investigate several strategies to avoid trapping of the $B$'s: moving an $A$ instead of the hindered $B$; allowing a trapped $B$ to hop with a small probability; biased walk towards non-hub nodes; and limiting the capacity of nodes. We obtain analytic results for lattices and networks, and discuss the advantages and shortcomings of the possible strategies.
\end{abstract}

\pacs{89.75.Hc,05.40.Fb,89.20.Hh}

\maketitle

\section{Introduction}
\label{intro}

The understanding of communication networks and the interplay between their structure and dynamics
has become an important research topic \cite{BA_review,PV_book,DM_book,Newman_review}.
In previous years, a variety of routing models have been proposed for the transport of messages over complex networks
using local \cite{WangfPhysRevE73,WangPhysRevE74,HuangChaos19,DeMartinoPhysRevE79,MeloniPhysRevE82,CH_book,Newman_book} and/or global characteristics of the underlying systems \cite{LingPhysRevE81} . These models are based on single species movement, attempting to improve the information flow efficiency.

In this article, we study the transport of messages in an environment where
they belong to different classes. In our model, two species, $A$ and $B$, diffuse independently, but where both species coexist only the high priority particles, $A$, are allowed to move. This problem
describes realistic scenarios in communication networks. In some networks, such as wireless sensor
networks \cite{RW_sensor1,RW_sensor2}, ad-hoc networks \cite{RW_adhoc1,RW_adhoc2} and peer-to-peer
networks \cite{RW_P2P}, data packets traverse the networks in a random fashion. Even when messages
are routed along shortest paths, in some networks the statistical properties of the traffic
resembles those of a random walk (see the appendix). Routers in communication networks often handle
both high and low priority information packets, such as, for example, in typical multimedia
applications. The low priority packets are sent out only after all high priority packets have been
processed \cite{KR_book,Tanenbaum_book}, just as in our model. [For a study of the jamming
transition under a dynamic routing protocol with priorities, see \cite{KimEPL2009}]. In the latter part of the article, we also consider some realistic extensions such as limited node capacity or a small probability for movement of a low priority message even in the presence of a high priority one.

We reported initial results for this model in \cite{PRERC}. We have shown that in lattices and
regular networks both species diffuse in the usual fashion, but the low priority $B$'s diffuse
slower than the $A$'s. In heterogeneous scale-free networks the $B$'s get mired in the high degree
nodes, effectively arresting their progress. Here we extend and generalize the main results of \cite{PRERC}. We then propose and analyze strategies to avoid the halting of the low priority messages, such as random walk models with soft priorities or with a bias, and discuss
their consequences in the context of communication networks. Our analytical results are summarized in Table \ref{SummaryTable}.

\section{Model definition}

In our model, whenever an $A$ or a $B$ particle is selected for motion, it hops to one of the
nearest neighbor sites with equal probability. We investigate two selection protocols. In the
\emph{site} protocol, a site is selected at random: if it contains both $A$ and $B$ particles, a
high-priority $A$ particle moves out of the site. A particle of type $B$ moves only if there are no
$A$'s on the site. If the site is empty, a new choice is made. In the \emph{particle} protocol, a
particle is randomly selected: if the particle is an $A$ it then hops out. A selected $B$ hops only
if there are no $A$ particles on its site. If the selected $B$ is not free, we consider two
subprotocols: \emph{(i)} \emph{'redraw'}: a new choice is made; \emph{(ii)} \emph{'moveA'}: One of
the coexisting $A$'s is moved instead of the $B$. With communication networks in mind, the site
protocol corresponds to selection of routers, whereas the particle protocol follows the trajectory
of individual data packets. Note that these protocols belong to the general framework of zero-range
processes, since the hopping rate of each particle depends only on the number of $A$'s and $B$'s in
its site (see, e.g., \cite{ZeroRange,ASEP} for factorized steady-state solutions and
\cite{Burda1,Burda2} for zero-range processes in networks). Our proposed scheme can also be described as a ``gas of particles'' model, where particle trajectories follow random walks.

As the underlying medium of the random walk, we consider lattices and two explicit random network
models: Erd\H{o}s-R\'{e}nyi (ER) random
networks \cite{ER,Bollobas}, where node degrees are narrowly (Poisson) distributed; and
scale-free (SF) networks, which are known to describe many communication networks and in particular
the Internet \cite{DIMES}. In SF networks, the degree distribution is broad, characterized by a
power-law tail $P(k)\sim k^{-\gamma}$, when usually $2<\gamma<3$
\cite{BA_review,PV_book,DM_book,Newman_review}. In the following sections, we will characterize the
diffusion of the high and low priority species in the different protocols and media.

\section{Analytical solution for lattices}

\label{sect_lattices}

\subsection{Number of empty sites}

We look first at lattices (or regular networks), where each site has exactly $z$ nearest neighbors.
While the $A$ particles move once they are selected, regardless of the $B$'s, the $B$'s, on the other
hand, can move only in those sites that are empty of $A$'s. Therefore, we begin by considering the
number of such sites, which we will later relate to the diffusion coefficients of the particles
under the priority constraints.

Define the number of sites as $N \rightarrow \infty$, and for now focus on a single species,
denoting its particle density by $\rho$. Let $f_j$ be the average equilibrium fraction of sites
that contain $j$ particles, and consider a Markov chain process whose states, $\{0,1,2,...\}$, are
the number of particles in a given site. The $\{f_j\}_{j=0,1,2,...}$ are the stationary
probabilities of the chain.

For the site protocol, the transition probabilities are:
\begin{equation}
\label{eq1} P_{j,j-1} = \frac{1}{N}\;;\quad P_{j,j+1} = \frac{1-f_0}{N}\;;
\end{equation}
$P_{j,j} = 1-P_{j,j-1}-P_{j,j+1}$ and all other transitions cannot occur. Indeed, for a site to
lose a particle it needs to be selected, with probability $\frac{1}{N}$. To gain a particle, one of
its $z$ non-empty neighbors must be chosen --- with probability $(1-f_0)\frac{z}{N}$ --- and this
neighbor must send the particle into the original site, with probability $\frac{1}{z}$. Note that
the final result is independent of the coordination number $z$, and is therefore independent on the
dimension and the lattice structure. Writing the master equations with the transition rates
(\ref{eq1}), we obtain:
\begin{equation}
\label{site_lattice} f_{j-1}(1-f_0)+f_{j+1}=f_j+f_j(1-f_0),
\end{equation}
and $f_1=(1-f_0)f_0$. This has the solution $f_j = f_0(1-f_0)^j$.
Imposing particle conservation, $\sum_{j=0}^{\infty}jf_j=\rho$, we
finally obtain:
\begin{equation}
\label{empty_site} f_0^{(\mbox{site})} = \frac{1}{1+\rho}.
\end{equation}

For the particle protocol the transition
probabilities are:
\begin{equation}
\label{transition_particle} P_{j,j-1} = \frac{j}{N\rho}\;;\quad P_{j,j+1} = \frac{1}{N}\;;
\end{equation}
$P_{j,j} = 1-P_{j,j-1}-P_{j,j+1}$ and all other transitions are excluded. Indeed, for a site to
lose a particle one of its $j$ particles (out of the total $N\rho$) needs to be selected. To gain a
particle, one of the $z\rho$ particles that reside, on average, in the neighboring sites has to be
chosen, and then hop to the original site (with probability $\frac{1}{z}$). Once again, the result
is independent of $z$. This time the boundary condition is $f_1=\rho f_0$, leading to
$f_j=f_0\frac{\rho^j}{j!}$. Imposing the normalization condition $\sum_{j=0}^{\infty}f_j=1$,
\begin{equation}
\label{empty_particle} f_0^{(\mbox{particle})} = e^{-\rho}.
\end{equation}
In other words, the $\{f_j\}$ are Poisson-distributed, with average $\rho$. This is expected,
having in mind that each of the total $N\rho$ particles is found in any of the lattice sites with
probability equal to $1/N$.

\subsection{Diffusion coefficients}
\label{theory_lattice}

We now employ the results of the previous subsection for the analysis of diffusion with priorities,
when \emph{both} species are involved. Assume that during the first $t$ steps of the protocol
(either one; counting successful steps only), $A$ has moved $n_A$ times and $B$ has moved $n_B$
times ($n_A+n_B=t$). The mean square displacement of the $A$ particles at time $t$ is:
\begin{equation}
\label{R2} \av{R_A^2(t)} = \av{\left[\sum_{i=1}^{n_A}\overrightarrow{r_i}\right]^2} =
\av{n_A}\av{r_i^2} = \av{n_A},
\end{equation}
since for the lattice, $\av{r_i^2}=1$. Denote $\av{n_A}=D_At$, such that $\av{R_A^2(t)}=D_At$.
Similar argument holds for the $B$ particles. Thus, both species diffuse as in the single-species
case, but due to the priority constraints, the time will be shared unevenly between the $A$'s and
$B$'s according to the diffusion coefficients to be found $D_A$ and $D_B$.

In the site protocol, a particle will surely move if we choose a non-empty site (containing $A$,
$B$, or both), which happens with probability $1-1/(1+\rho_A+\rho_B)$. This is true, because the
particles behave as a single, non-interacting species if one ignores their labeling, and thus Eq.
\eqref{empty_site} can be invoked with $\rho=\rho_A+\rho_B$. An $A$ particle moves if the selected
site contains any number of $A$'s, which happens with probability $\rho_A/(1+\rho_A)$, again, from
Eq. \eqref{empty_site}. Therefore, $D_A=\frac{\rho_A}{1+\rho_A}/(1-\frac{1}{1+\rho_A+\rho_B})$, or:
\begin{equation}
\label{P_A} D_A = \frac{\rho_A(1+\rho_S)}{(1+\rho_A)\rho_S}\;;\quad D_B =
\frac{\rho_B}{(1+\rho_A)\rho_S}\,,
\end{equation}
where $\rho_S\equiv\rho_A+\rho_B$ and we have used $D_B=1-D_A$ for
the second relation. Simulation results for the site protocol confirming Eq. \eqref{P_A} were presented in \cite{PRERC}.

For the particle protocol, denote the ratio of free $B$ particles (i.e., $B$ particles not sharing
their site with $A$'s) to all $B$ particles by $r$. In the \emph{redraw} subprotocol, no particle
will move when a non-free $B$ particle is chosen, which happens with
probability $\rho_B/\rho_S\cdot (1-r)$. In the \emph{moveA} subprotocol, any particle will always
move (since a non-free $B$ gives its turn to an $A$). In both subprotocols, a $B$ particle moves
whenever a free $B$ is chosen, with probability $\rho_B/\rho_S\cdot r$. Therefore:
\begin{equation}
\label{P_B} D_B^{(\rm{redraw})} = \frac{r\rho_B/\rho_S}{1-(1-r)\rho_B/\rho_S}\;;\quad
D_B^{(\rm{moveA})} = r\rho_B/\rho_S\,,
\end{equation}
and $D_A=1-D_B$.

\begin{figure}[ht]
\centering \includegraphics[width=1.\linewidth] {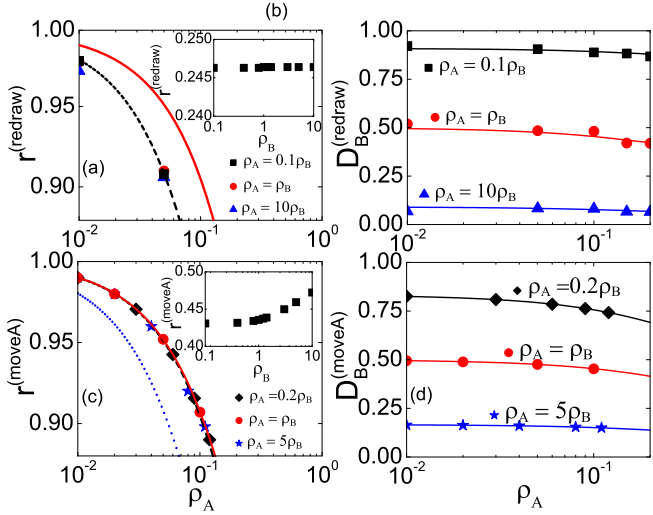}
\caption{(Color online) Priority diffusion model on lattices. (a) The probability of a $B$ particle to be free for the \emph{redraw} subprotocol, $r^{(redraw)}$, vs $\rho_A$. The dashed black line is for Eq. \eqref{r_second_order} and the red solid line is for $\exp(-\rho_A)$. Inset: $r^{(redraw)}$ vs. $\rho_B$. (b) The diffusion coefficient of the low priority $B$ particles for the \emph{redraw} subprotocol, $D_B^{(redraw)}$, vs $\rho_A$. Solid lines are for Eq. \eqref{P_B} after substituting Eq. \eqref{r_second_order}. (c), (d) Same as (a) and (b), respectively, for the \emph{moveA} subprotocol. In (c), the dotted blue line is for Eq. \eqref{r_second_order} for the \emph{moveA} subprotocol, the solid red line is for $\exp(-\rho_A)$, and the dashed black line is, for comparison, for Eq. \eqref{r_second_order} for the \emph{redraw} subprotocol.}
\label{Figure1_lattice}
\end{figure}

Had the density of $B$'s been independent of the $A$'s, then $r$ would simply be the fraction of
sites empty of $A$, or $r=e^{-\rho_A}$. However, due to the priority constraints, the concentration of the $B$'s is not uniform. In Figure \ref{Figure1_lattice}(a) and Figure \ref{Figure1_lattice}(c), we present simulation results for $r$. For the
\emph{redraw} subprotocol, the $B$'s tend to stick with the $A$'s, such that $r \lesssim
e^{-\rho_A}$. For the \emph{moveA} protocol, the $B$'s tend to repel from the $A$'s (since once an
$A$ enters a site that has $B$'s, whenever \emph{any} particle will be chosen, the $A$ will be forced out), so that $r \gtrsim e^{-\rho_A}$ (see the slight difference at the lower part of Figure \ref{Figure1_lattice}(c)). While we are not able to solve for $r$ in the general case, the low density regime is amenable for a direct solution that displays all the above mentioned features.

For low densities, we make the approximation that a single site cannot contain more than one $A$ or one $B$. We use again a Markov chain formulation, but now with just four possible states to each
site: $\{\phi,A,B,AB\}$ (state $A$ corresponds to a site having one $A$ particle, and similarly for the other states). We write the transition probabilities as before, to first order in the
densities:
\begin{eqnarray}
\label{MC_smalld}
\nonumber && P_{\phi,A} = \frac{\rho_A}{N\rho_S} \;;\quad
P_{\phi,B} = \frac{\rho_B}{N\rho_S} \;;
\\ \nonumber &&
P_{A,\phi} = \frac{1}{N\rho_S} \;;\quad
P_{A,AB} = \frac{\rho_B}{N\rho_S}  \;;
\\ \nonumber &&
P_{B,\phi} = \frac{1}{N\rho_S} \;;\quad
P_{B,AB} = \frac{\rho_A}{N\rho_S} \;;
\\ &&
P_{AB,B}^{(\rm{redraw})} = \frac{1}{N\rho_S}\;;\quad
P_{AB,B}^{(\rm{moveA})} = \frac{2}{N\rho_S}.
\end{eqnarray}
Unindicated transition probabilities are zero, and the diagonal accounts for normalization
$P_{x,x}=1-\sum_{y \ne x}P_{x,y}$. The justification is similar to that of
Eq.~(\ref{transition_particle}). For a site to lose a particle, this particle needs to be chosen
out of a total of $N\rho_S$ particles. For a site to gain an $A$, one of the $z\rho_A$ particles
that reside, on average, in the neighboring sites has to be chosen (out of $N\rho_S$), and then
sent to the target site, with probability $\frac{1}{z}$. The probability to gain a $B$ is similar,
since in the first-order approximation we ignore non-free $B$'s. The priority constraint is taken
into account by forbidding the transition $AB \rightarrow A$. $AB$ is transformed into $B$ whenever the $A$ is chosen (for the \emph{redraw} subprotocol), or whenever either the $A$ or the $B$ is chosen (for the \emph{moveA} subprotocol).

From the master equations of the chain we solve for $r \equiv \frac{f_B}{f_B+f_{AB}}$ to first
order:
\begin{align}
\label{r_first_order} &r^{(\rm{redraw})} = 1-2\rho_A+{\cal O}(\rho^2)\;;\nonumber \\
&r^{(\rm{moveA})} = 1-\rho_A+{\cal O}(\rho^2)
\end{align}
($\rho$ stands for either $\rho_A$ or $\rho_B$). To obtain the next order, allowed states can have
two particles of each type ($\{\phi,A,AA,B,AB,AAB,BB,ABB,AABB\}$), and we take into account that
when a $B$ is chosen it actually hops only with probability $r$ (using its first-order expression,
Eq.~(\ref{r_first_order})). This gives
\begin{align}
\label{r_second_order} &r^{(\rm{redraw})} = 1-2\rho_A+\frac{13}{4}\rho_A^2 + {\cal
O}(\rho^3)\;;\nonumber \\&r^{(\rm{moveA})} = 1-\rho_A+\frac{3}{4}\rho_A^2 + {\cal O}(\rho^3).
\end{align}
The prediction for $r$, as well as the diffusion coefficients obtained on substituting Eq.~(\ref{r_second_order}) in (\ref{P_B}), compares well with simulations (Figure \ref{Figure1_lattice}). From Eq. \eqref{r_second_order}, it can be seen that $r$ does not depend on $\rho_B$, at least to second order (for both subprotocols). In fact, our
simulations suggest that for the \emph{redraw} subprotocol $r$ is independent of $\rho_B$ for all
densities (inset of Figure \ref{Figure1_lattice}(a)). Intuitively, this happens because the probability for a $B$ to be free is dictated by the presence of $A$ particles and not by other $B$ particles. In contrast, in the \emph{moveA} subprotocol $r$ is increasing with $\rho_B$ as the probability of an $A$ particle to be pushed out of a site increases with increasing $\rho_B$ (inset of Figure \ref{Figure1_lattice}(c)). Comparing the expansion of $r$ in the two subprotocols with $e^{-\rho_A}=1-\rho_A+\frac{1}{2}\rho_A^2+{\cal O}(\rho^3)$ we find $r^{(\rm{redraw})} \lesssim e^{-\rho_A} \lesssim r^{(\rm{moveA})}$, as expected.

\section{Priority diffusion in networks}

\label{networks_sect}

We now turn to heterogeneous networks, where the degree $k$ varies from site to site. We focus on
the particle protocol, and later discuss briefly the site protocol, which yields qualitatively
similar results. We start with the fraction of empty sites \emph{of degree} $k$,
$f_0^{(k)}$. Consider a network with only one particle species and define a Markov chain on the
states $\{0,1,2,...\}$ for the number of particles in a given site of degree $k$. The stationary
probabilities are $f_j^{(k)}$. The chain has the transition probabilities:
\begin{equation}
P_{j,j-1} = \frac{j}{N\rho}\;;\quad P_{j,j+1} = \frac{k}{\av{k}}\frac{1}{N}\;;
\end{equation}
$P_{j,j} = 1-P_{j,j-1}-P_{j,j+1}$ and all other probabilities are zero. $P_{j,j-1}$ is same as in
Eq.~(\ref{transition_particle}). For a site to gain a particle, a neighboring site must first be
chosen, and there are $k$ such sites. Since the neighbor is arrived at by following a random link,
the probability that it has degree $k'$ is $k'P(k')/\av{k}$ (see, e.g., \cite{Generating}), and in
that case, it will have, on average, $\rho k'/\av{k}$ particles (see below or, e.g., \cite{Rieger}). Since the particle is sent back to the original site with probability $1/k'$, the overall
probability for the original site to gain a particle is:
\begin{equation}
\label{particles_in_neighbor}
k\sum_{k'=1}^{\infty}\left[\frac{k'P(k')}{\av{k}} \cdot
\frac{\frac{\rho k'}{\av{k}}}{N\rho} \cdot \frac{1}{k'}\right] =
\frac{k}{\av{k}}\frac{1}{N}.
\end{equation}
Solving for the stationary probabilities while keeping in mind that $\sum_j f_j^{(k)}=1$, one finds
\begin{equation}
\label{empty_k} f_j^{(k)} = f_0^{(k)}\frac{\left(\rho k/\av{k}\right)^j}{j!}\;;\quad f_0^{(k)} =
\exp\left({-\rho k/\av{k}}\right).
\end{equation}
Note that for regular networks, when all sites have the same degree, this reduces to Eq.
\eqref{empty_particle}, $f_0=e^{-\rho}$. The average number of particles in a site of degree $k$ is
$\sum_{j=0}^{\infty}jf_j^{(k)}=\rho k/\av{k}$, as expected.

When both species are involved, consider the \emph{redraw} subprotocol where the $A$'s move
independently of the $B$'s, and define that in one time step each particle has on average one
moving attempt. At each time step, a $B$ particle in a node of degree $k$ has, on average,
probability $\exp(-\rho_A k/\av{k})$ to jump out (Eq. \eqref{empty_k}), since this is the
probability of that site to have no $A$'s (assuming that the interaction between species is weak, as in lattices for large $\rho_A$ \cite{PRERC}). This results in an exponential distribution of waiting
times (for a $B$ particle):
\begin{equation}
\label{eq_psi_t_vs_k}
\psi_k(t) = \frac{1}{\tau_k}e^{-t/\tau_k},
\end{equation}
with $\tau_k\equiv\exp(\rho_A k/\av{k})$. Simulation results confirming Eqs. \eqref{empty_k} and \eqref{eq_psi_t_vs_k} were shown in \cite{PRERC}.

\begin{figure}[ht]
\centering
\subfigure[]{
   \includegraphics[width=8cm] {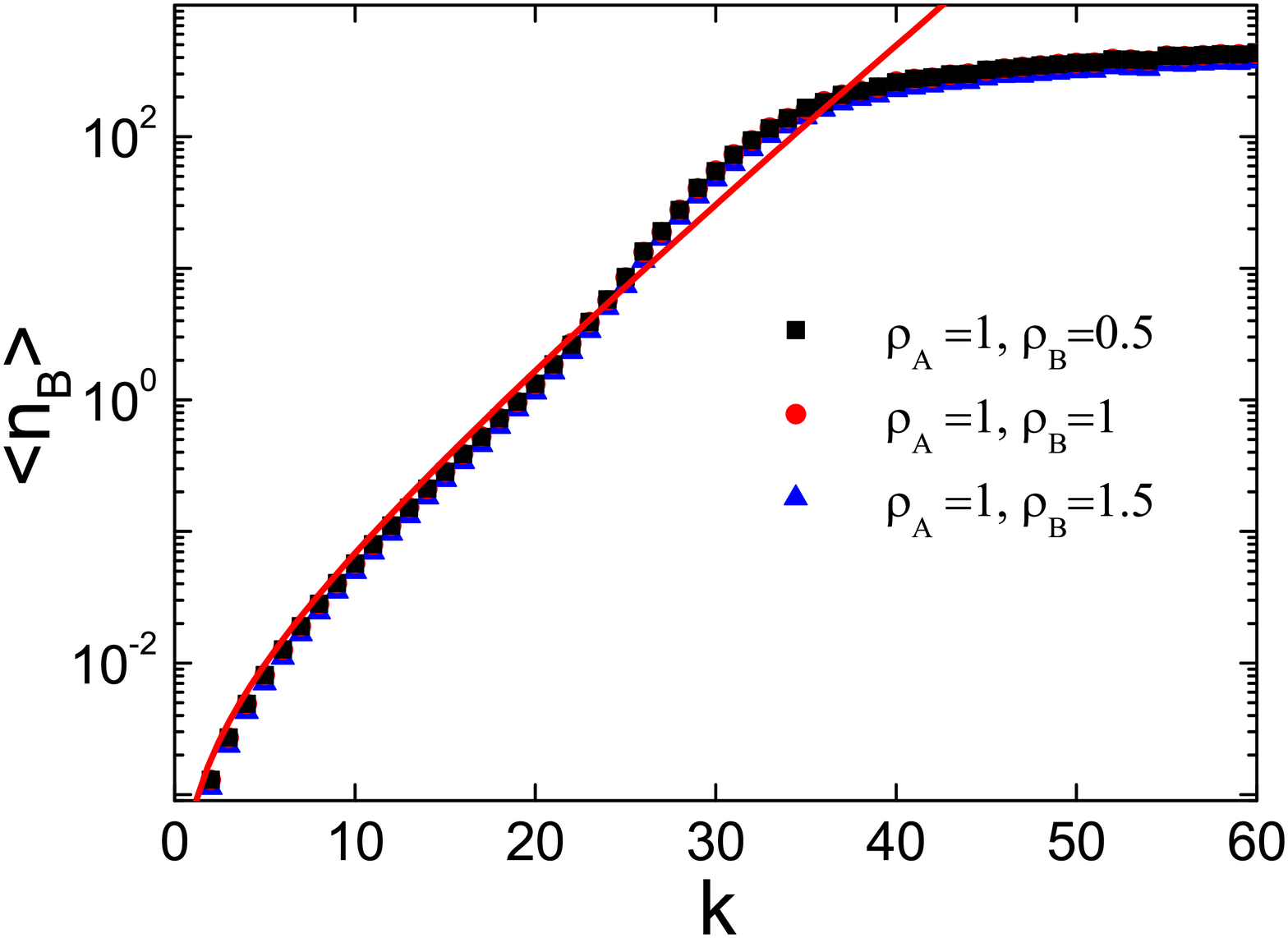}
   \label{Figure2_a_SFN_redraw_particle}
}
\subfigure[]{
   \includegraphics[width=8cm] {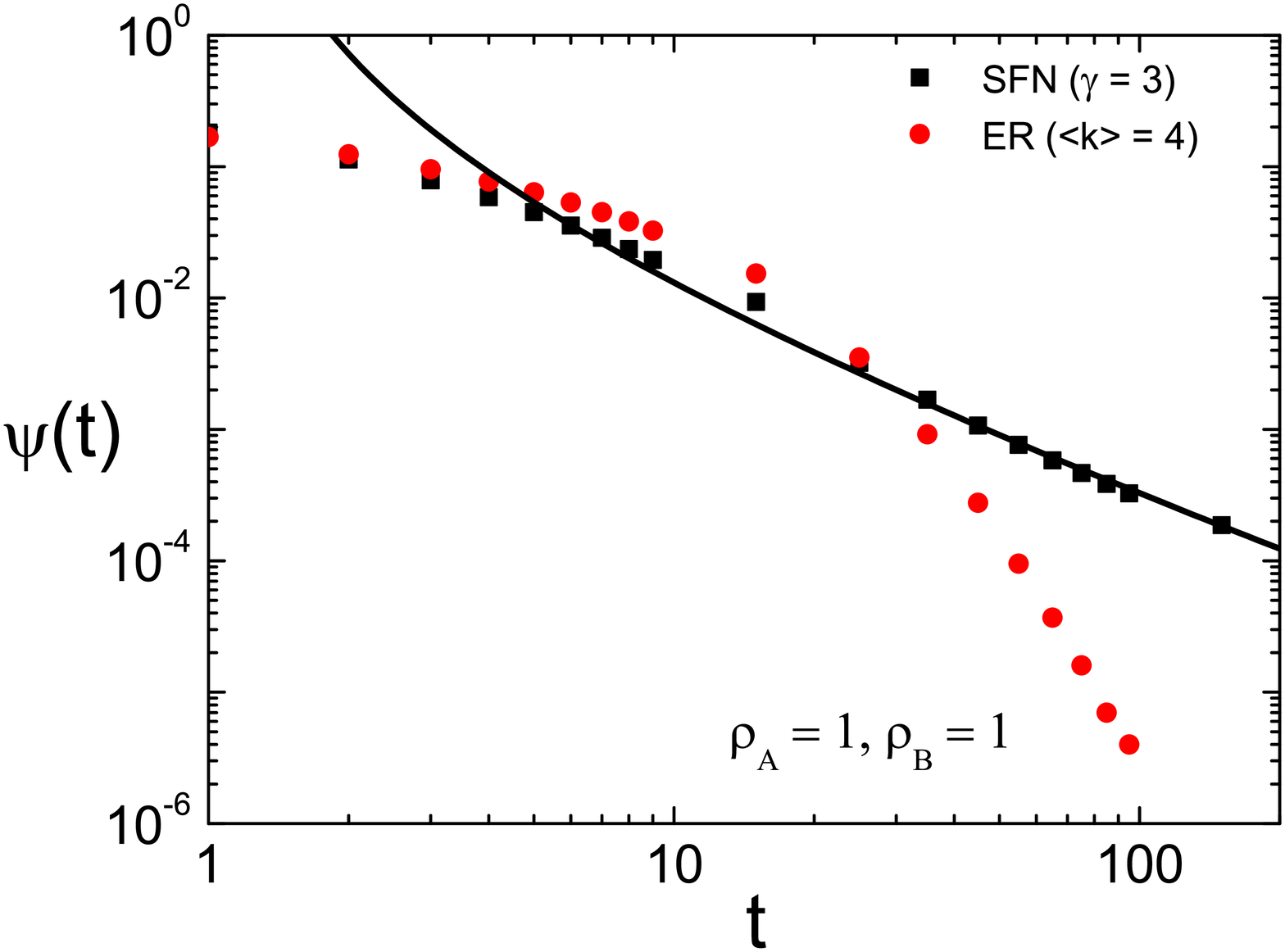}
   \label{Figure2_c_SFN_redraw_particle}
}
\caption{(Color online) The priority diffusion model on scale-free networks. (a) The average concentration of low priority particles, $n_B$, vs. the node degree $k$ for the redraw subprotocol in case of scale-free networks ($\gamma=3$, $k\geq 2$; generated as in \cite{MR}; symbols). Curves for different values of $\rho_B$ were normalized to collapse. The solid line stands for $\av{n_B}\sim k\exp(\rho_A k/\av{k})$. (b) The distribution of waiting times $\psi(t)$ for $\rho_A=\rho_B=1$ for the case of the \emph{redraw} subprotocol in scale-free networks (black squares) and ER networks with $\av{k}=4$ (red circles). The black solid line represents Eq. \eqref{psi}, $\psi(t)\sim\frac{1}{\ln^{\gamma-1}t}$, further divided by $t$ to account for the fact that during our simulations, waiting times are sampled with probability inversely proportional to their duration (e.g., short waiting times are sampled more than long waiting times).}
\label{Figure2_lattice}
\end{figure}

The exponentially long waiting time (in the degree $k$) means that in heterogeneous networks such
as scale-free networks --- where the degrees may span several orders of magnitude --- the $B$
particles get mired in the hubs (high degree nodes). The problem is exacerbated by the fact that
the $B$ particles are drawn to the hubs even in the absence of $A$'s: the presence of $A$'s only
amplifies this tendency. In fact, the concentration of the $B$'s is proportional to $k\exp(\rho_A
k/\av{k})$. To see this, denote by $n^{(B)}_i$ the number of $B$'s at node $i$. The probability of
a $B$ to hop from node $i$ to a neighboring node $j$ in one time step is $p_B(i,j) = \exp(-\rho_A
k_i/\av{k})\frac{1}{k_i}$, the product of the probability that node $i$ is free of $A$'s ( $\exp(-\rho_A
k_i/\av{k})$) and the probability the particle is sent to node $j$ ($1/k_i$).
In equilibrium, the number of $B$'s getting in and out of a node should be equal:
$\sum_{j}n^{(B)}_ip_B(i,j) = \sum_{j}n^{(B)}_jp_B(j,i)$, and these equations are satisfied by
$n^{(B)}_i \propto k_i\exp(\rho_A k_i/\av{k})$ (with the prefactor calculated from $\sum_kP(k)n^{(B)}(k)=\rho_B$). This is confirmed in Figure \ref{Figure2_a_SFN_redraw_particle}. Therefore, in large scale-free networks the $B$'s collect at the hubs and their diffusion is
effectively halted.

Using $n^{(B)}\sim k\exp(\rho_A k/\av{k})$, the probability of a random particle to reside in a
node of degree $k$ is $G(k)\sim k\exp(\rho_A k/\av{k})P(k)$, where $P(k)$ is the degree
distribution. We can now use $G(k)$ to find the waiting time distribution of a random particle,
$\psi(t)$. Since $\psi_k(t)$ is relatively narrow, we replace it by a delta function $\psi_k(t) =
\delta(t-\tau_k)$, or $t(k) = \exp(\rho_A k/\av{k})$. For SF networks where $P(k)\sim k^{-\gamma}$,
changing variables $\psi(t)dt=G(k)dk$ gives
\begin{equation}
\label{psi} \psi(t) \sim \frac{1}{\ln^{\gamma-1}t}.
\end{equation}
The waiting time distribution is
therefore broad, with some particles stalling for very long times. Eq. \eqref{psi} for the waiting time distribution in SF networks is compared to simulations in Figure \ref{Figure2_c_SFN_redraw_particle}, as well as to the much narrower distribution in ER networks. Eq. \eqref{psi} is expected to hold only up to time $\exp(\rho_Ak_{\textrm{max}}/\av{k})$, where
$k_{\text{max}} \sim N^b$, with $b=1/2$ for $2 < \gamma < 3$ and $b=\frac{1}{\gamma-1}$ for $\gamma\geq 3$ \cite{Boguna,BurdaZ}.

\begin{figure}[ht]
\centering
\subfigure[]{
   \includegraphics[width=8cm] {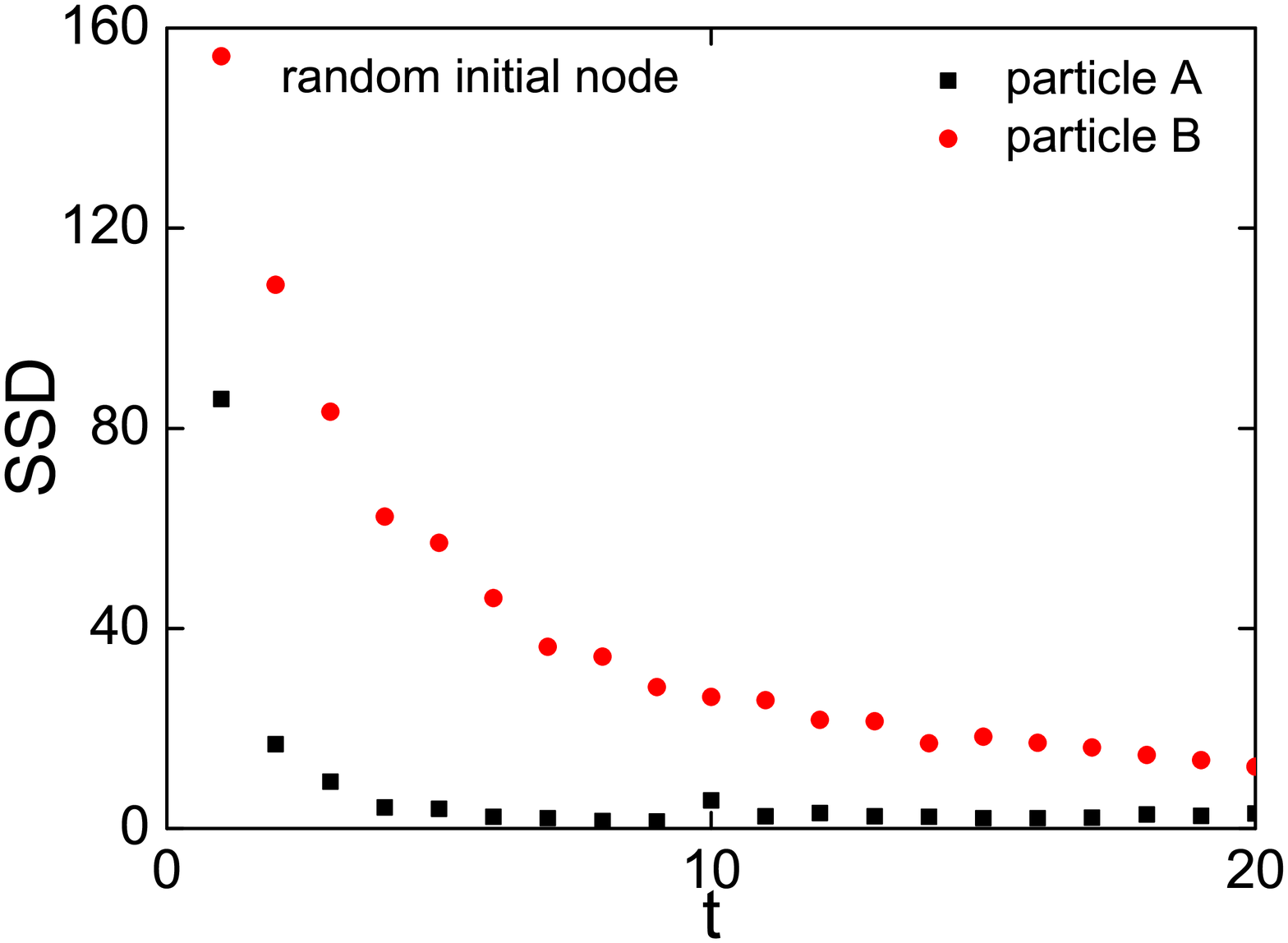}
   \label{Figure_dynamics_a}
}
\subfigure[]{
   \includegraphics[width=8cm] {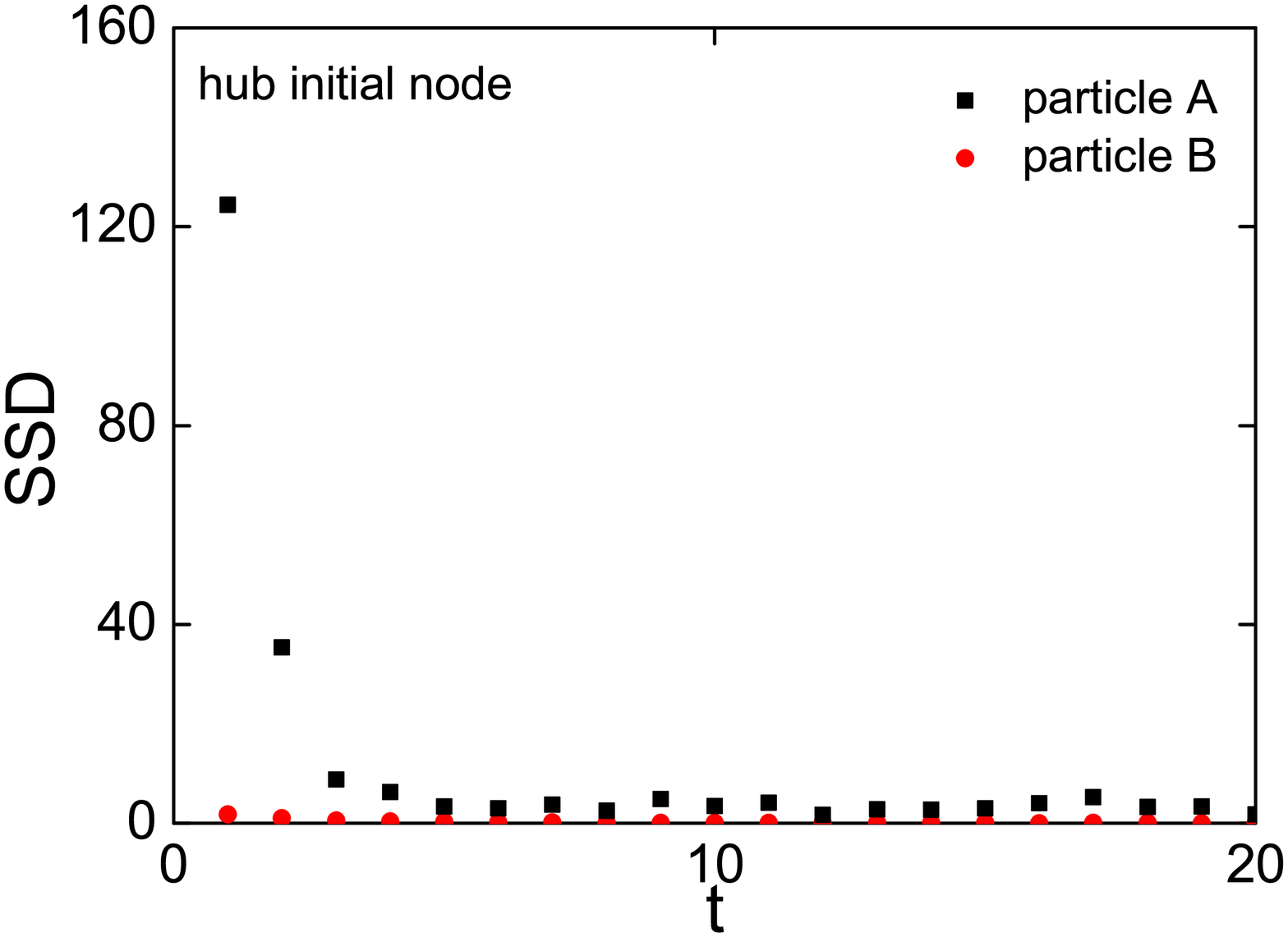}
   \label{Figure_dynamics_b}
}
\caption{(Color online) The dynamics of approach to equilibrium in the priority diffusion model. In each panel, the Sum of Squared Differences ($\textrm{SSD}$) between the concentration profiles $\av{n(k)}$ at successive time steps is plotted for both particle species. (a) $A$ and $B$ particles are randomly assigned an initial node. (b) All $A$ particles are initially placed in the largest hub and all $B$ particles are placed in the second largest hub. The overall concentrations were $\rho_A=1$ and $\rho_B=1$.}
\label{Figure_dynamics}
\end{figure}

Analytical and simulation results have so far have assumed that the system is in equilibrium. Specifically, in each simulation, we used a ``burn-in'' period of $2000$ Monte Carlo steps. To investigate the dynamics of reaching equilibrium, we examined, in Figure \ref{Figure_dynamics}, the rate at which the concentration profile $n(k)$ approaches its equilibrium form. This was quantified as
the Sum of Squared Differences (SSD) between the profiles at consecutive time points:
\begin{equation}
\label{ssd} \textrm{SSD}(t)= \sum_{k=2}^{k_m}{(\av{n_{t}(k)}-\av{n_{t-1}(k)})^2},
\end{equation}
where $\av{n_{t}(k)}$ is the average particle concentration (either $A$'s or $B$'s) at nodes of degree $k$ at time $t$ and we set $k_m=50$. The results are shown in Figure \ref{Figure_dynamics} for two classes of initial conditions: (\emph{i}) $A$'s and $B$'s are randomly distributed over all nodes (Figure \ref{Figure_dynamics_a}) and (\emph{ii}) all $A$'s are placed in the largest hub and all $B$'s are placed in the second largest hub (Figure \ref{Figure_dynamics_b}). Uniform distribution has been tested and produces similar results as in Figure \ref{Figure_dynamics_a}.  In both cases, both species of particles reach equilibrium rapidly; but interestingly, $B$ particles equilibrate slower than $A$'s for uniform initial conditions and faster when initially placed on the hub. This happens because in equilibrium, most $B$'s are at the hubs, and therefore, if they start at the hub they will tend to remain in place. However, if the $B$'s are initially uniformly distributed, the priority constraints will slow them down on their way to reaching the hubs.

\begin{figure}[ht]
\centering \includegraphics[width=8cm] {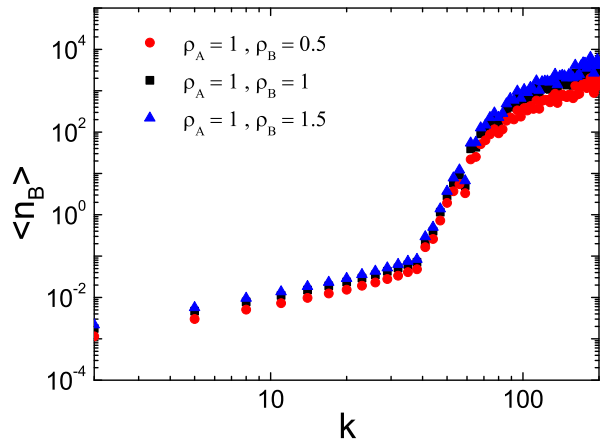}
\caption{(Color online) The site protocol on scale-free networks. Plotted is the average concentration of the low priority particles, $n_B$, vs. the node degree $k$ for the site protocol on scale-free networks ($\gamma=3$, $k\geq 2$). The concentration increases with the node degree in a complex manner involving two phases (for which we have no theoretical arguments); the location of the transition point is in fact time-dependent (not shown).}
\label{Figure3_a_SFN_redraw_site}
\end{figure}

For the site protocol, the analytical approach presented in this section is not directly applicable, because transition probabilities for different degrees cannot be decoupled. Intuitively,
however, it is clear that also for the site protocol the number of particles increases with the
site degree. To see this, consider again a single species, and assume that the concentration $\rho$ is large enough that selected sites are never empty. For a given site, the probability to lose a
particle is $1/N$ (the probability of the site to be selected). The probability to gain a particle
is $\frac{1}{N}\sum_{j}\frac{1}{k_j}$, where the sum is over all $k$ neighbors of the site. Since
the latter term scales as $k$, the number of gained particles is expected to increase with the site
degree. With two species, that would again imply trapping of the low
priority particles, just as in the particle protocol. This behavior is demonstrated in
Figure \ref{Figure3_a_SFN_redraw_site}.

To summarize so far, the combination of \emph{(i)} heterogenous network structure, \emph{(ii)} random walk, and \emph{(iii)} strict priority policy leads to slowing down of the low priority particles. In the next section, we investigate strategies to enhance the mobility of the $B$ particles even under priority constraints.

\section{Strategies to avoid trapping of $B$'s}

Given a heterogenous network structure, how can one implement a random walk with priorities, and yet guarantee the low priority particles are not halted?

\subsection{\emph{moveA} subprotocol}

\begin{figure}[ht]
\centering
\subfigure[]{
   \includegraphics[width=8cm] {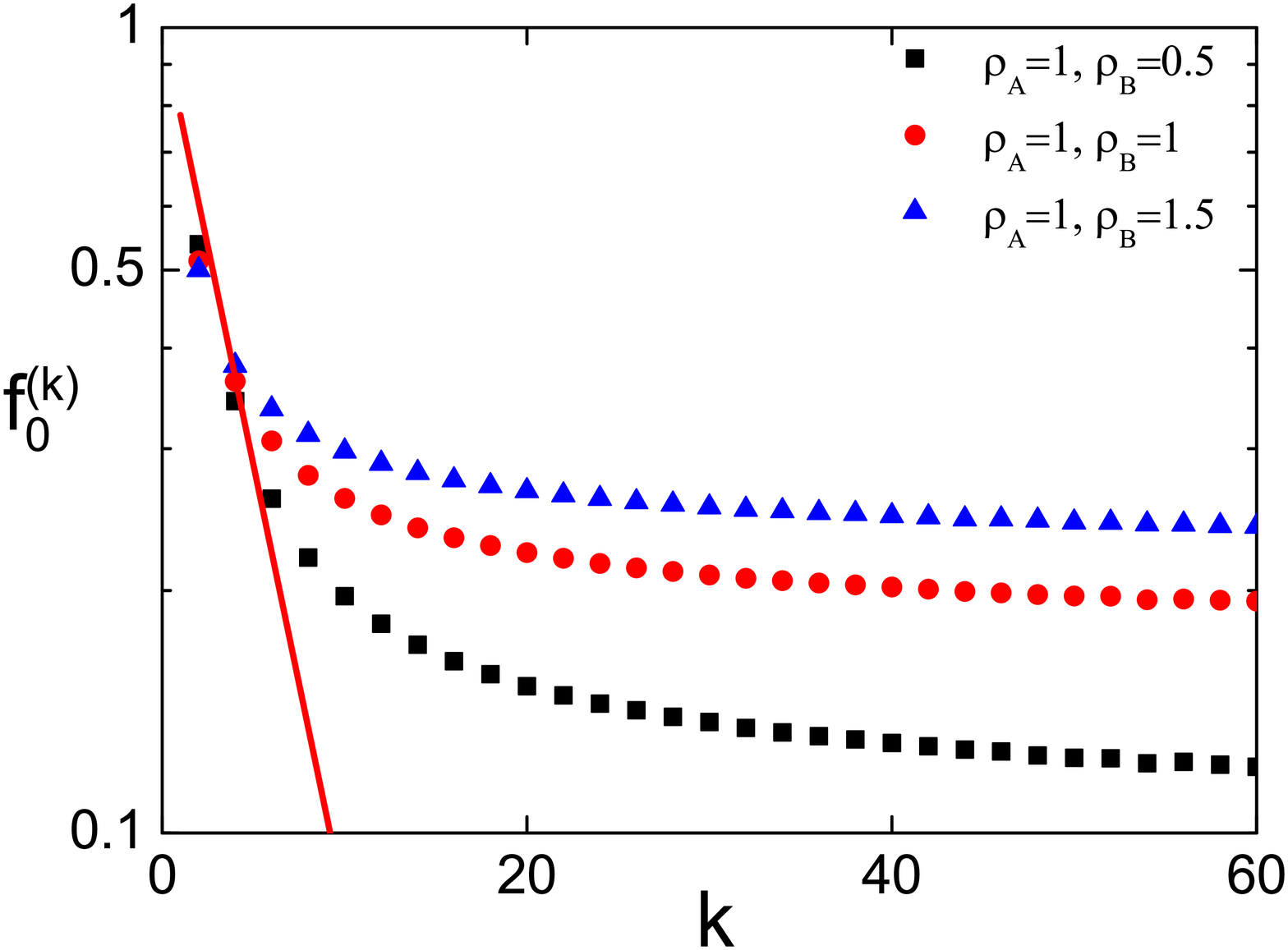}
   \label{figure4_freeSitesOfA_moveA_SFN}
}
\subfigure[]{
   \includegraphics[width=8cm] {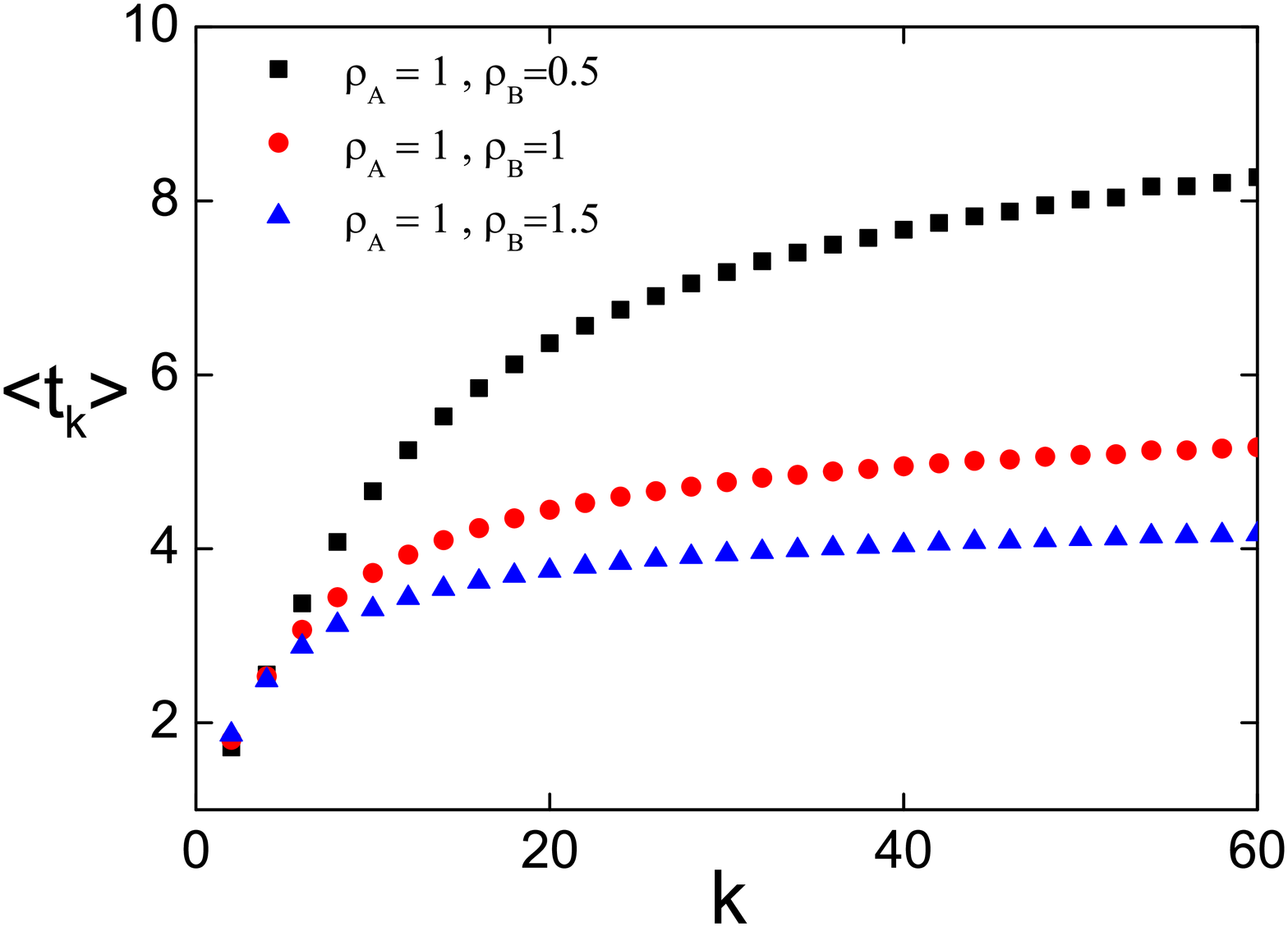}
   \label{figure4_meanTime_moveA_SFN}
}
\caption{(Color online) The \emph{moveA} protocol in networks. (a) The fraction of sites empty of $A$ particles, $f_0^{(k)}$, vs. the node degree $k$. The solid red line stands for $\exp(-\rho_A k/\av{k})$, the expression for the \emph{redraw} protocol. (b) The average waiting time $\av{t_k}$ vs. the node degree $k$. For large $k$, the waiting time is close to a constant.}
\label{Figure4_lattice}
\end{figure}

Recall the \emph{moveA} subprotocol, in which $A$'s mobility is driven both by a selection of $A$'s and by a selection of arrested $B$'s. For this subprotocol, there is no trapping of the $B$'s since $A$'s do not aggregate at the hubs, but are rather rejected from them. Once an $A$ arrives into a node with many $B$'s, there is high probability for a $B$ to be chosen and push the $A$ outside the node. For this subprotocol, we numerically show that the probability of a site to be empty of $A$'s decays slower than exponentially in $k$, and that the average waiting time for a $B$ is short and almost independent of the degree (Figure \ref{Figure4_lattice}).

\subsection{Soft priorities}

Consider a \emph{soft} priority model, in which a $B$, when co-localized with $A$'s, has a small
probability $\epsilon$ of leaving the site. As we show below, this results in enhanced diffusion of the $B$'s, even for networks.

\begin{figure}[ht]
\centering
\subfigure[]{
   \includegraphics[width=8cm] {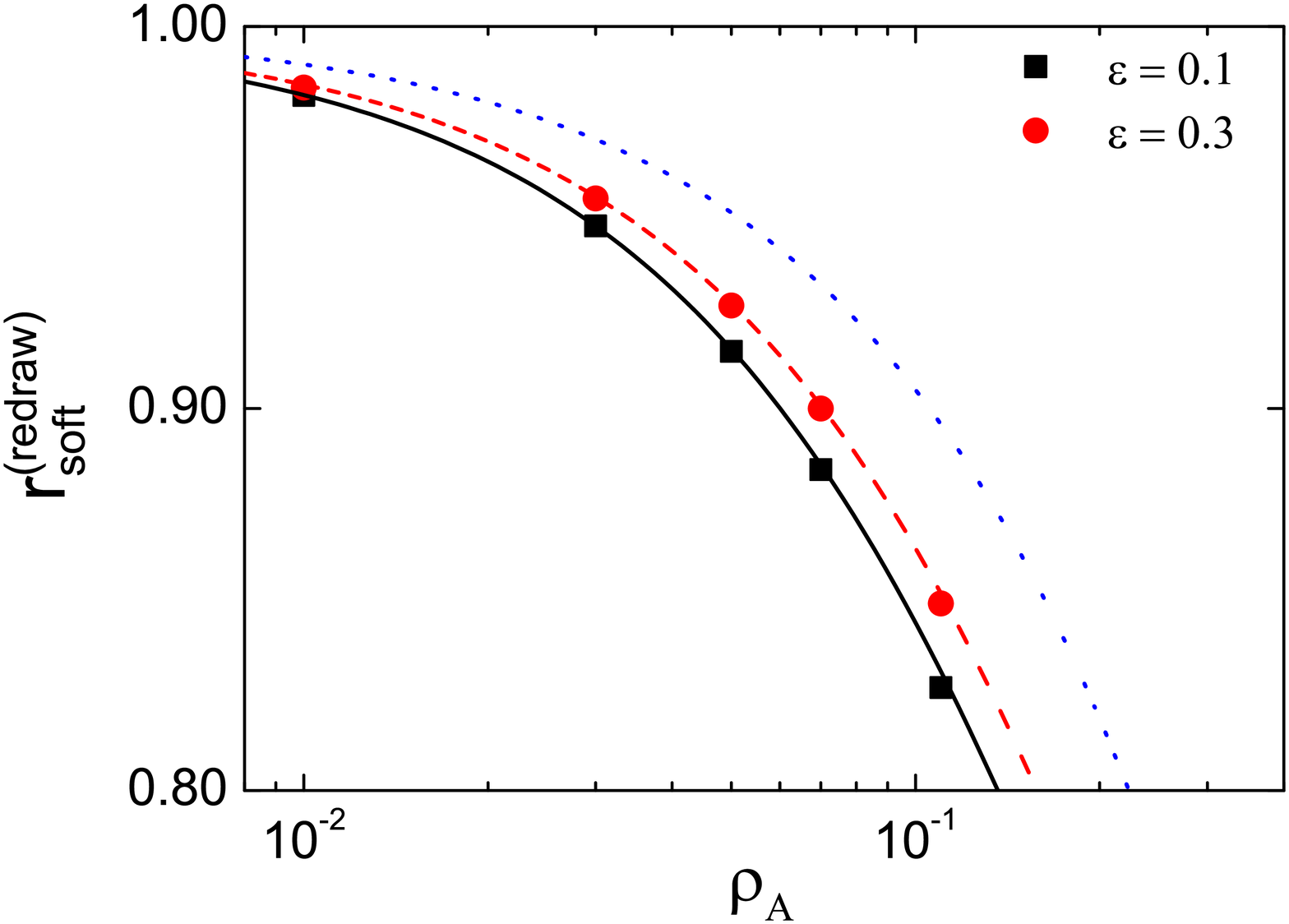}
   \label{figure5_r_redraw_psoft_lattice}
}
\subfigure[]{
   \includegraphics[width=8cm] {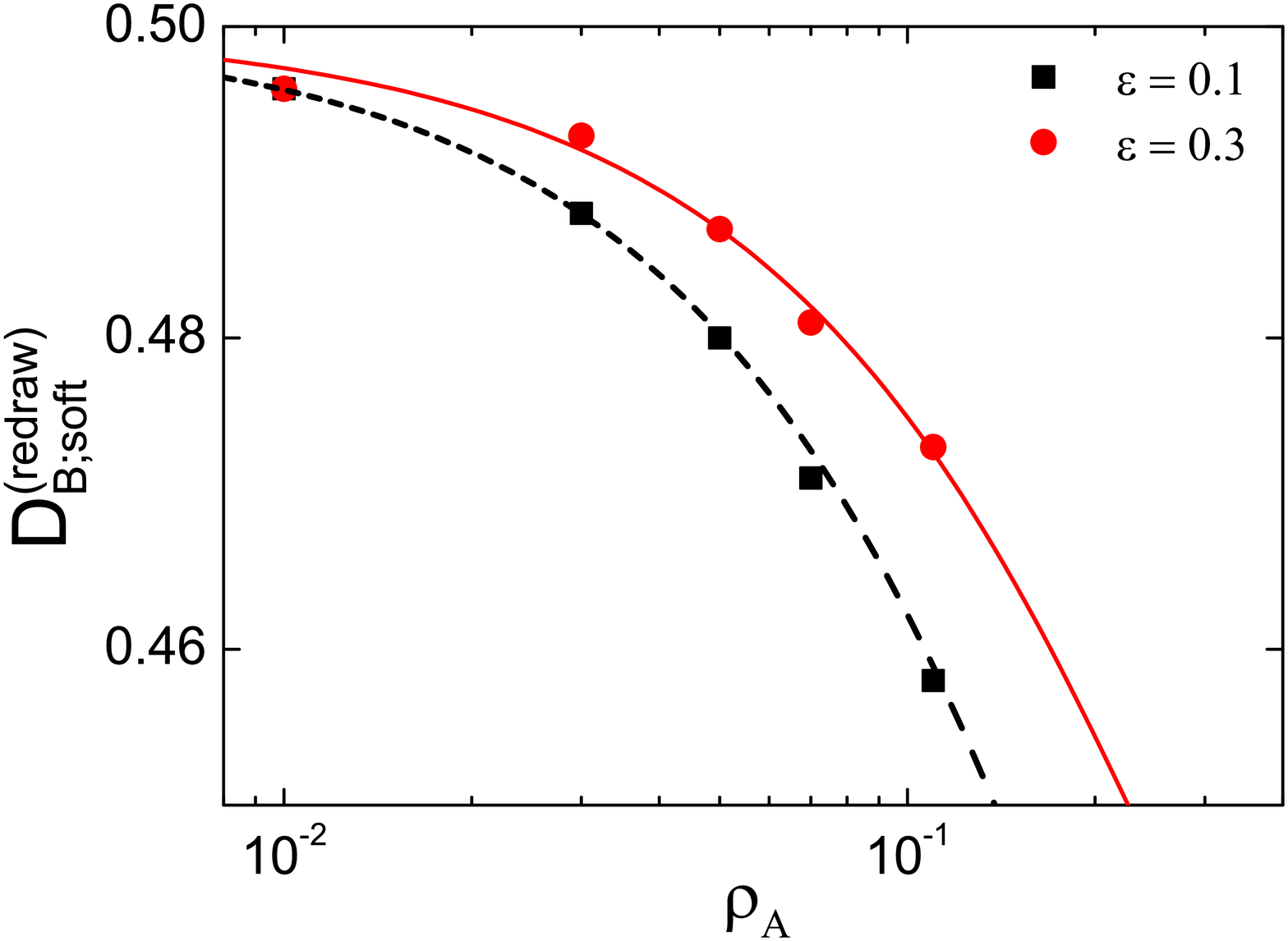}
   \label{figure5_DB_redraw_psoft_lattice}
}
\caption{(Color online) Soft priorities on lattices. (a) The probability of a $B$ particle to be free in lattices with soft priorities, $r_{\textrm{soft}}^{(\textrm{redraw})}$, vs. $\rho_A$. The $B$ particle density is $\rho_B=\rho_A$. Solid black and dashed red lines are for Eq. \eqref{r_second_order_soft}; the dotted blue line stands for $\exp(-\rho_A)$. (b) The diffusion coefficient of the $B$ particles, $D_{B;\textrm{soft}}^{(\textrm{redraw})}$, vs. $\rho_A$. Lines are for Eq. \eqref{PB_soft}.}
\label{Figure5_lattice}
\end{figure}

Consider lattices first. An analytical solution can be derived as in the strict priority model
(Section \ref{theory_lattice}), if the last line of Eq. (\ref{MC_smalld}) becomes:
\begin{eqnarray}
\nonumber && P_{AB,B}^{(\rm{redraw})} = \frac{1}{N\rho_S}\;;\quad P_{AB,B}^{(\rm{moveA})} =
\frac{2-\epsilon}{N\rho_S} \;;\
\\ &&
P_{AB,A}^{(\rm{redraw})} = \frac{\epsilon}{N\rho_S}\;;\quad P_{AB,A}^{(\rm{moveA})} =
\frac{\epsilon}{N\rho_S}.
\end{eqnarray}
Using the last equation, the fraction of free $B$'s, up to first order, is:
\begin{align}
&r^{(\rm{redraw})} = 1-\frac{2}{1+\epsilon}\rho_A+{\cal O}(\rho^2)\;;\nonumber \\
&r^{(\rm{moveA})} = 1-\rho_A+{\cal O}(\rho^2).
\end{align}
As in Section \ref{theory_lattice}, the first order solution can be substituted in the equations
for the larger Markov chain that allows for two particles of the same species in a single site.
Here, a $B$ will be free to move with probability $r+(1-r)\epsilon$. Solving the second order
problem, we find:
\begin{align}
\label{r_second_order_soft} & r^{(\rm{redraw})} =
1-\frac{2}{1+\epsilon}\rho_A+\frac{13+2\epsilon-3\epsilon^2}{2(1+\epsilon)^2(2+\epsilon)}\rho_A^2 + {\cal O}(\rho^3)\;;\nonumber\\
&r^{(\rm{moveA})} = 1-\rho_A+\frac{3-\epsilon}{4}\rho_A^2 + {\cal O}(\rho^3).
\end{align}
Note that Eq. \eqref{r_second_order_soft} reduces to Eq. \eqref{r_second_order} in the case
$\epsilon=0$, and to the series expansion of $e^{-\rho_A}$ for $\epsilon=1$ (no priorities, $A$'s
and $B$'s are independent). Eq. \eqref{r_second_order_soft} is compared to simulations in Figure
\ref{figure5_r_redraw_psoft_lattice}. As for the case of strict priorities, $r$ is independent of $\rho_B$ and approaches $e^{-\rho_A}$ for large densities. The diffusion coefficients are:
\begin{align}
\label{PB_soft} & D_B^{(\rm{redraw})} = \frac{\left[r +
(1-r)\epsilon\right]\rho_B/\rho_S}{1-(1-r)(1-\epsilon)\rho_B/\rho_S}\;;\nonumber\\
&D_B^{(\rm{moveA})} = \left[r + (1-r)\epsilon\right]\rho_B/\rho_S.
\end{align}
Eq. \eqref{PB_soft} is compared to simulations in Figure \ref{figure5_DB_redraw_psoft_lattice}. As expected, the diffusion of
the $B$'s is always accelerated whenever $\epsilon>0$ for the \emph{redraw} subprotocol. For the
\emph{moveA} subprotocol, $r$ decreases with increasing $\epsilon$, since $A$'s are rejected from
sites that have $B$'s less often than in the $\epsilon=0$ case. However, at least for small $\rho_A$, the diffusion coefficient for the $B$'s increases with $\epsilon$: for small $\rho_A$, $r^{(\rm{moveA})}$ is very weakly dependent of $\epsilon$ (Eq. \eqref{r_second_order_soft}), while $D_B^{(\rm{moveA})}$ increases linearly with $\epsilon$ (Eq. \eqref{PB_soft}).

\begin{figure}[ht]
\centering
\subfigure[]{
   \includegraphics[width=8cm] {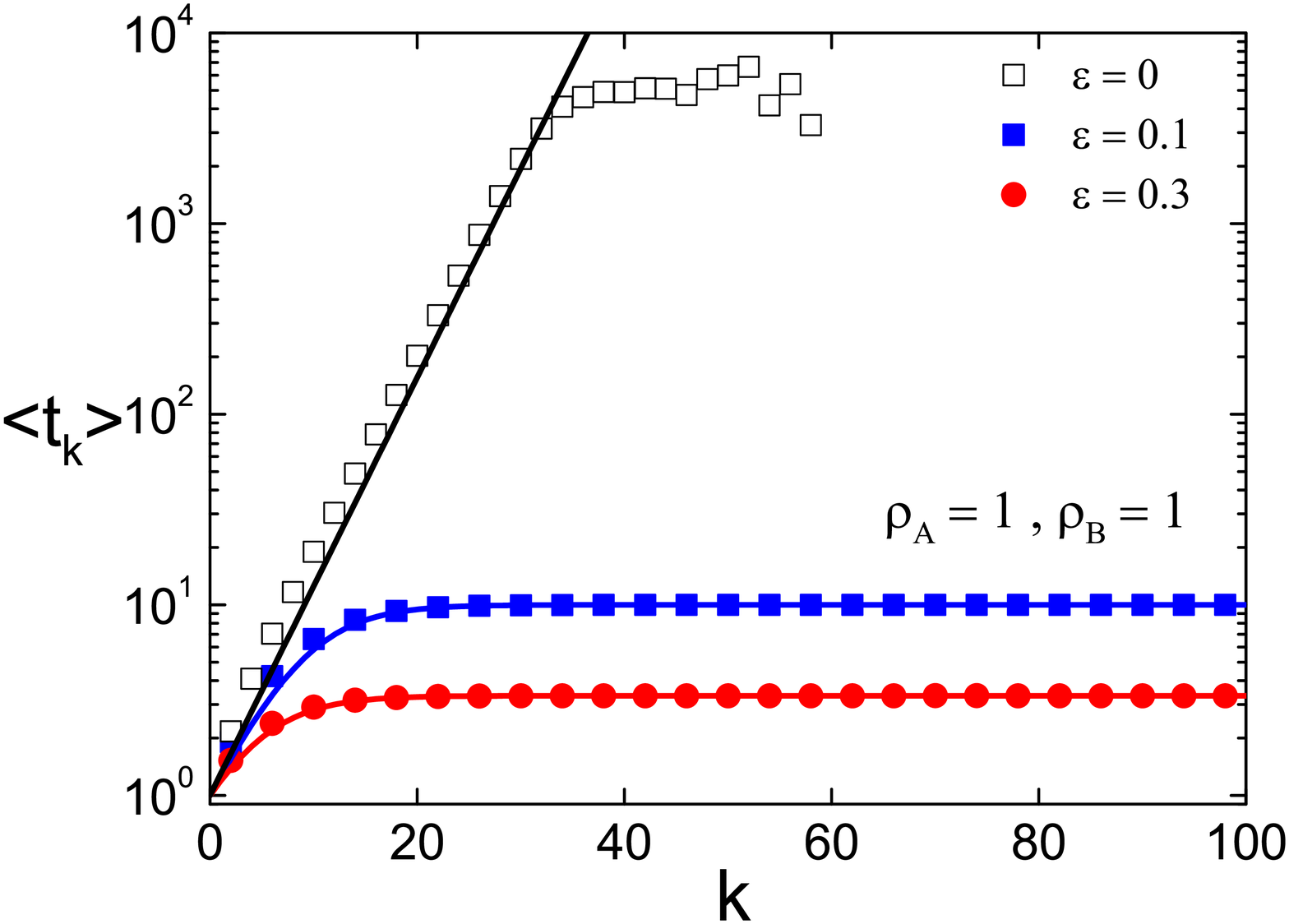}
   \label{figure6_meanTime_psoft_SFN}
}
\subfigure[]{
   \includegraphics[width=8cm] {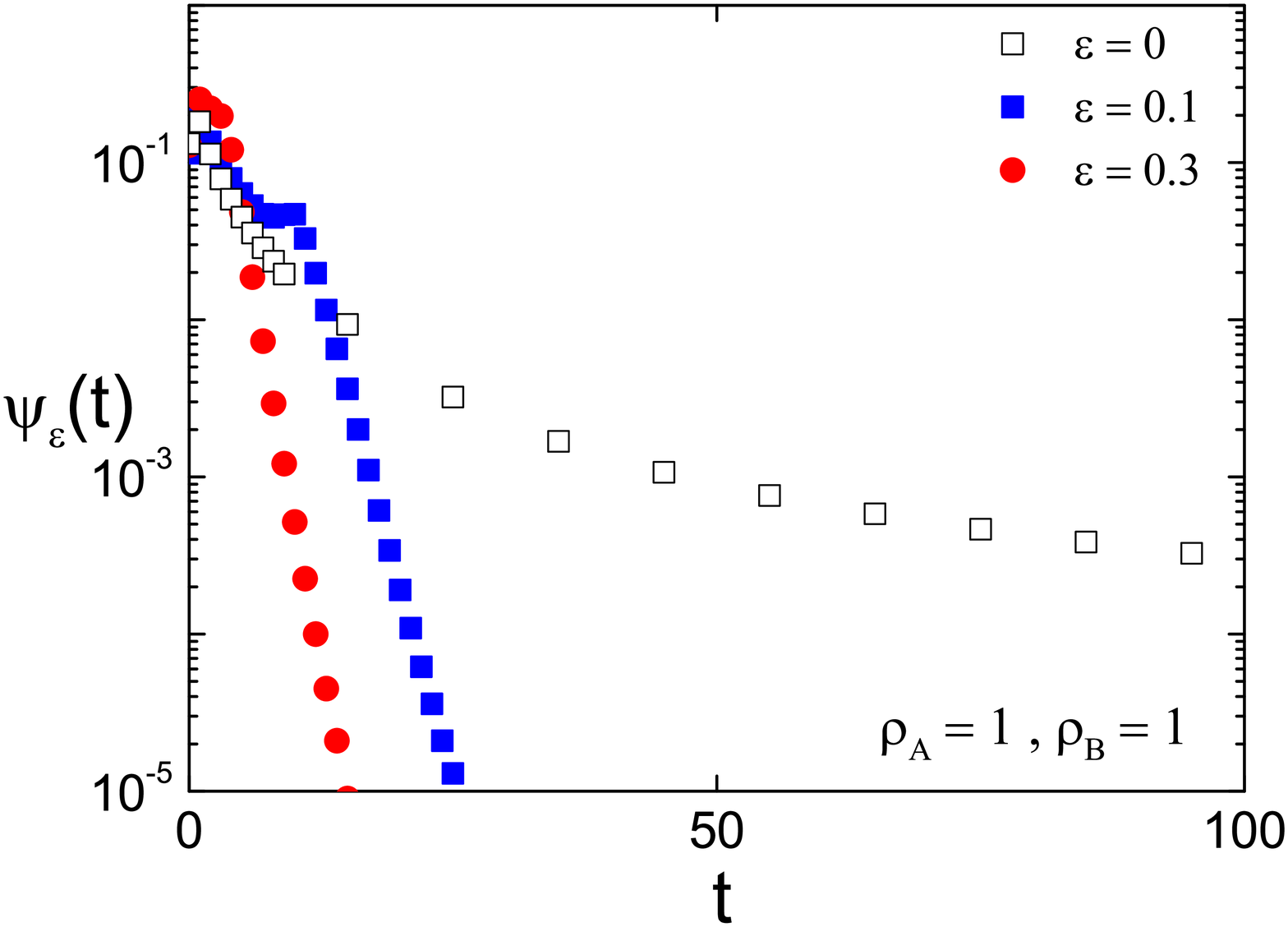}
   \label{figure6_psiEpsilon_psoft_SFN}
}
\caption{(Color online) Soft priorities on networks. (a) The average waiting time of $B$ particles, $\av{t_k}$, vs. $k$, for the case of soft priorities \emph{redraw} subprotocol (scale-free networks with $\gamma=3$ and $k\geq 2$; $\rho_A = \rho_B = 1$). Solid lines are for Eq. \eqref{tau_x}. (b) The distribution of waiting times $\psi_{\epsilon}(t)$ for $\epsilon = 0$ (open squares), $\epsilon = 0.1$ (blue squares), and $\epsilon = 0.3$ (red dots). }
\label{Figure6_SFN}
\end{figure}

For networks, consider the particle protocol in the \emph{redraw} version. $B$ particles can move if either
(\emph{i}) they are free, with probability $\exp(-\rho_A k/\av{k})$ or (\emph{ii}) if they coexist with an $A$ but are allowed to jump, with probability $\epsilon\left[1-\exp(-\rho_A k/\av{k})\right]$. The
average waiting time is the inverse of the hopping probability:
\begin{equation}
\label{tau_x} \tau_k = \frac{1}{\exp(-\rho_A k/\av{k})+\epsilon[1-\exp(-\rho_A k/\av{k})]}.
\end{equation}
From Eq. (\ref{tau_x}), even for $k \rightarrow \infty$, $\tau_k \sim \epsilon^{-1}$ and thus
diverges with $k$ only for $\epsilon=0$. This is confirmed in simulations (Figure \ref{figure6_meanTime_psoft_SFN}).
Therefore, even the slightest escape probability is sufficient to avoid the trapping of the low
priority particles. In Figure \ref{figure6_psiEpsilon_psoft_SFN}, we plot the distribution of the low priority particles
waiting times, $\psi_{\epsilon}(t)$, for different values of $\epsilon$. As expected, for $t\gg \epsilon^{-1}$ the decay is exponential.

\subsection{Avoiding hubs}

\label{sect_avoidhubs}

\begin{figure}[ht]
\centering
\subfigure[]{
   \includegraphics[width=8cm] {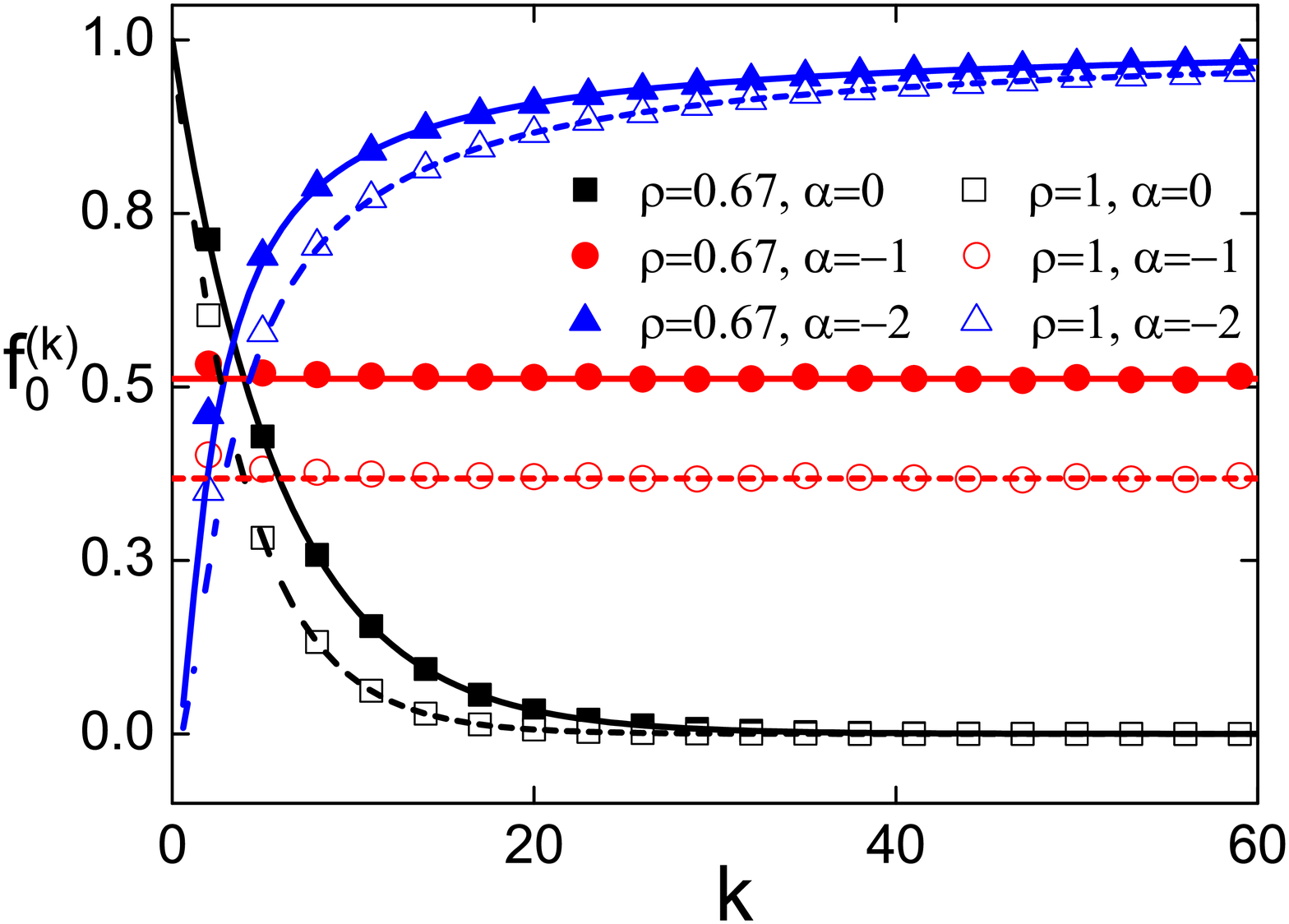}
   \label{figure7_freeSites_oneParticle_SFN_avoidHubs}
}
\subfigure[]{
   \includegraphics[width=8cm] {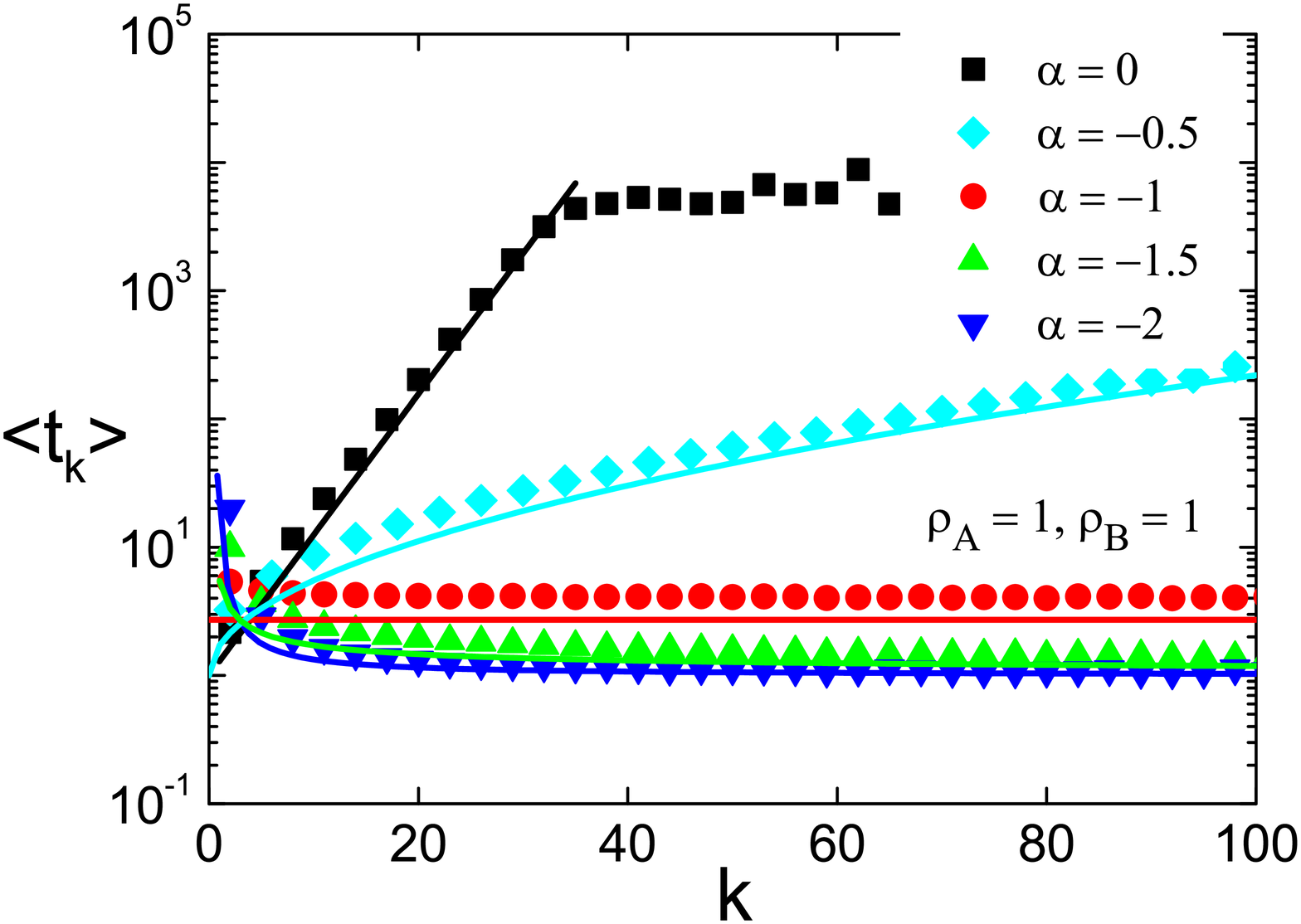}
   \label{figure7_meanTime_avoidHubs_PDM_SFN}
}
\caption{(Color online) The \emph{avoid hubs} protocol. (a) The fraction of empty sites for one species, $f_0^{(k)}$, vs. the node degree $k$. Lines stand for Eq. \eqref{avoiding_hubs_f0}. (b) The average waiting time of $B$ particles, $\av{t_k}$, vs. $k$. Solid lines correspond to $\av{t_k}=1/f_0^{(k)}=\exp\left(\rho k^{1+\alpha}/\av{k^{1+\alpha}}\right)$.}
\label{figure7_avoidHubs_PDM_SFN}
\end{figure}

One of the necessary conditions leading to the trapping of the $B$'s is the tendency of random
walkers to concentrate at the hubs. This can be restrained if one assumes that each node is
familiar with the degrees of its neighbors, and can thus avoid high degree nodes whenever possible. Consider the following model, where particles choose their next step according to the following
rule \cite{Agata}:
\begin{equation}
\label{bias_rule} P_{ij} = \frac{k_j^{\alpha}}{\sum_{m} k_m^{\alpha}},
\end{equation}
where $j$ is a neighbor of $i$ and the sum runs over all neighbors. Writing again a Markov chain
for the number of particles per site (for a single species), the transition probabilities are:
\begin{equation}
P_{j,j-1} = \frac{j}{N\rho}\;;\quad P_{j,j+1} = \frac{k^{1+\alpha}}{\av{k^{1+\alpha}}}\frac{1}{N}.
\end{equation}
The probability to gain a particle was calculated as follows. Each neighboring node of the given
site has probability $k'P(k')/\av{k}$ to have degree $k'$, and has on average $\rho
k'^{1+\alpha}/\av{k^{1+\alpha}}$ particles \cite{Agata}. The neighbor sends the particle to the
given site with probability $k^{\alpha}/\frac{k'\av{k^{1+\alpha}}}{\av{k}}$ \cite{Agata}. Thus, the
probability to gain a particle is:
\begin{equation}
k\sum_{k'=1}^{\infty}\left[\frac{k'P(k')}{\av{k}} \cdot
\frac{\frac{\rho k'^{1+\alpha}}{\av{k^{1+\alpha}}}}{N\rho} \cdot
\frac{k^{\alpha}}{\frac{k'\av{k^{1+\alpha}}}{\av{k}}}\right] =
\frac{1}{N}\frac{k^{1+\alpha}}{\av{k^{1+\alpha}}}.
\end{equation}
Following the same steps as before, this leads to:
\begin{equation}
\label{avoiding_hubs_f0} f_0^{(k)} = \exp\left({-\frac{\rho
k^{1+\alpha}}{\av{k^{1+\alpha}}}}\right).
\end{equation}
Eq. \eqref{avoiding_hubs_f0} is compared to simulations in Figure \ref{figure7_freeSites_oneParticle_SFN_avoidHubs}. For $\alpha=-1$,
$f_0^{(k)}$ is independent of $k$ and we recover the lattice case. Whenever $\alpha>-1$, particles
tend to aggregate at the hubs as before, leading again to trapping of low priority particles. When
$\alpha<-1$, particles are attracted to the small nodes. However, since there are many small nodes, this does not lead to any further halting of the $B$'s. This is demonstrated in Figure
\ref{figure7_meanTime_avoidHubs_PDM_SFN}, where the average waiting time of the $B$'s is plotted vs. $\alpha$. Requiring nodes to be aware of their neighbors' degrees is reasonable in the context of communication networks, since
this information can be attached to messages or exchanged between the nodes. Similar ideas were
developed in \cite{bottleneck} in the context of routing in communication networks. The drawback of the method is that by avoiding the hubs, it takes the particles more time to cover the
network.

\subsection{Limited nodes capacity}

\begin{figure}[ht]
\centering
\subfigure[]{
   \includegraphics[width=8cm] {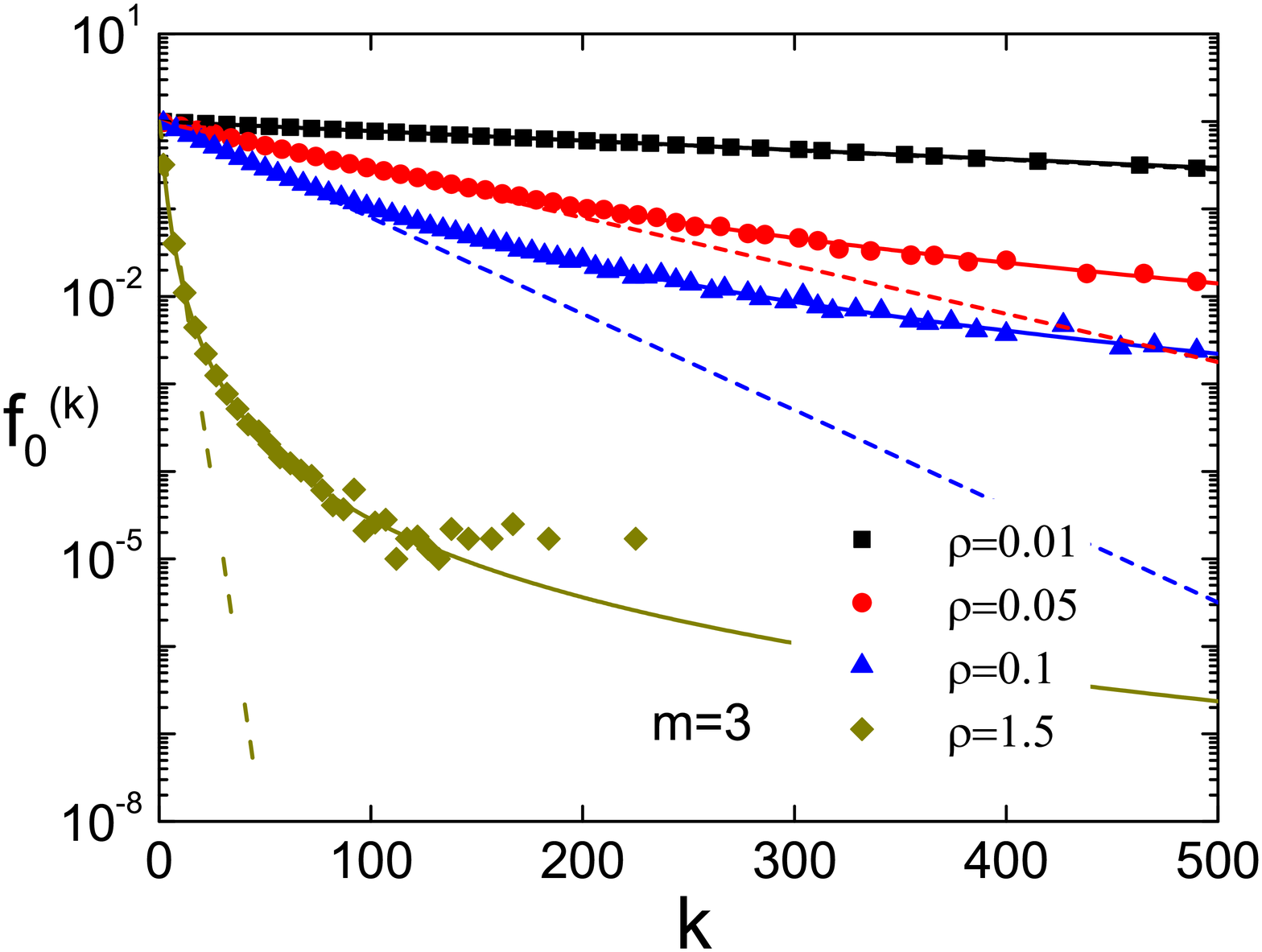}
   \label{figure8_freeSites_ConcCapacity3_1particle_SFN}
}
\subfigure[]{
   \includegraphics[width=8cm] {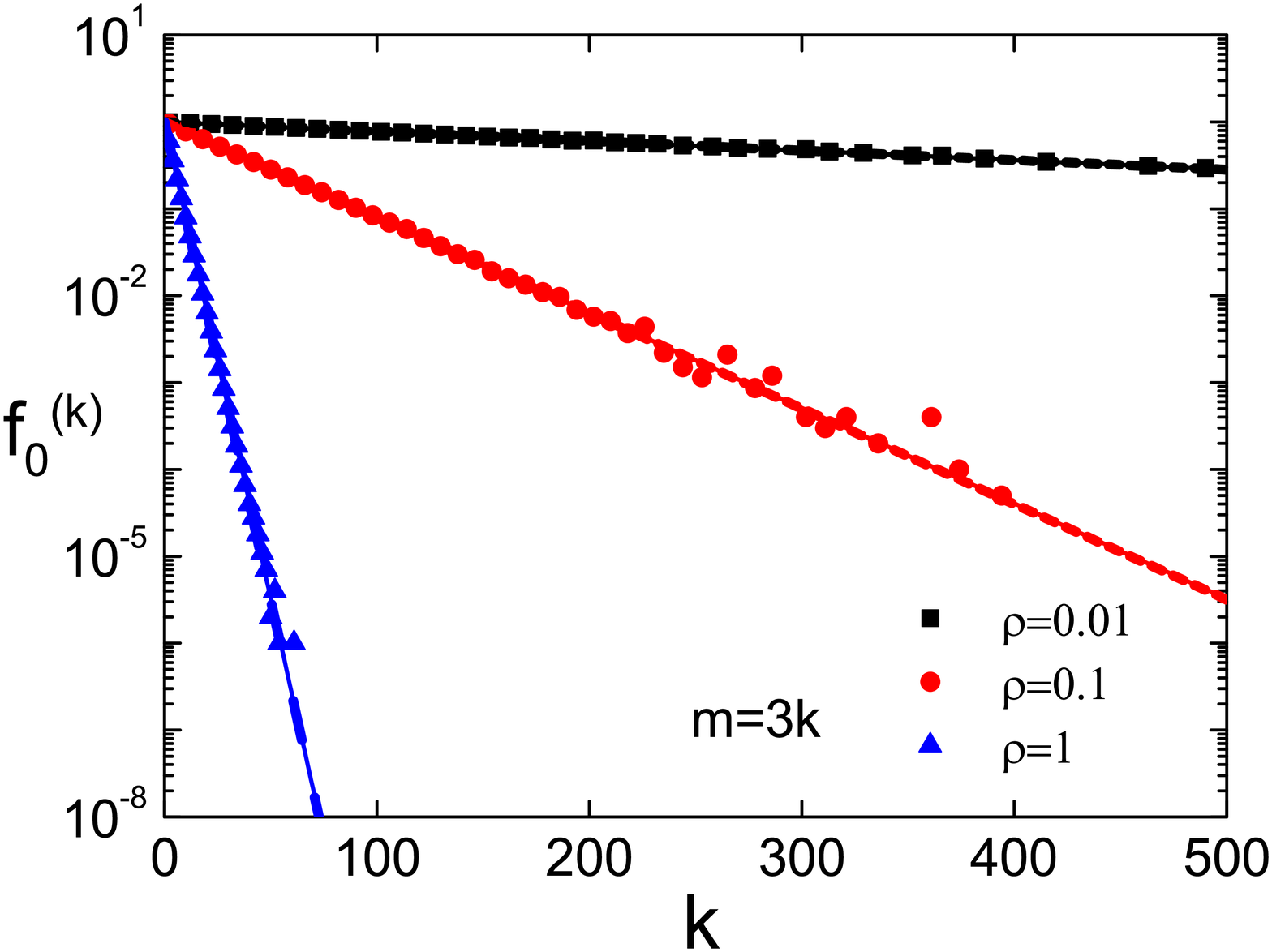}
   \label{figure8_freeSites_varCapacity3_1particle_SFN}
}
\caption{(Color online) Limited node capacity for one species. The fraction of empty sites for one species, $f_0^{(k)}$, vs. the node degree $k$ for (a) constant capacity $m=3$ and (b) variable capacity $m(k)=3k$. Full symbols present simulation results. Solid lines correspond to Eq. \eqref{fj_limited_capacity}, while dashed lines correspond to the infinite capacity case, $f_0=\exp(-\rho k/\av{k})$.}
\label{figure8_freeSites_Capacity_1particle_SFN}
\end{figure}

In real communication applications, routers may be able to store only a limited amount of
information. In our model, this constraint would translate to a limited number of particles in a
node, such that particles cannot jump into nodes that have reached their capacity \cite{Capacity,Chiense2011}. The analysis of such a model is complicated by the fact that particles are interacting even for a single species; for example, when the capacity is one particle per node, the particles are effectively fermions \cite{deMoura2005,SandersPRE2009}. We could nevertheless find an approximation to the fraction of empty sites. For concreteness, assume that each node has capacity $m(k)$ and that the single-species particle density is $\rho$. At each time step, a particle, selected at random, attempts to jump into one of its neighbors. However, if that neighbor is full, the jump is unsuccessful and the particle remains in place. Using again the Markov chain for the number of particles per site, and similarly to \cite{SandersPRE2009}, the transition probabilities for a node of degree $k$ are
\begin{align}
\label{limited_capacity_probs}
&P_{j,j-1} = \frac{j}{N\rho}\sum_{k'=1}^{\infty}\frac{k'P(k')}{\av{k}}\left(1-f_m^{(k')}\right)\equiv \frac{jC}{N\rho} \\ &P_{j,j+1} =
k\sum_{k'=1}^{\infty}\frac{k'P(k')}{\av{k}}\frac{n(k')}{N\rho}\frac{1}{k'}=\frac{k}{N\av{k}}~~(j<m) \nonumber.
\end{align}
The probability to lose a particle is $j/(N\rho)$, but then multiplied by the probability that the neighbor site is not full, $(1-f_m)$. Since $f_m$ could be different for different degrees, we need to condition on the neighbor's degree, but as the sum does not depend on either $k$ or $j$, it is a constant that depends on $\rho$ and $\av{k}$ only and will be found later by computing the average density. In the second line, the probability to gain a particle is as in Sections \ref{networks_sect} and \ref{sect_avoidhubs}, except that we denote the average number of particles in a site of degree $k'$ as $n(k')$. Using the relation $\sum_{k}P(k)n(k)=\rho$, the final transition probability is in fact as in the unconstrained Xcase.
Using Eq. \eqref{limited_capacity_probs} and the normalization condition, it can be shown that the stationary probabilities satisfy (for $j=0,1,...,m(k)$)
\begin{equation}
\label{fj_limited_capacity}
f_j^{(k)}=\frac{\left(\frac{\rho k}{C\av{k}}\right)^j/j!}{\sum_{j'=0}^{m(k)}\left(\frac{\rho k}{C\av{k}}\right)^{j'}/j'!}.
\end{equation}
The constant $C$ is found by solving
\begin{equation}
\label{C_limited_capacity}
\sum_{k=1}^{\infty}P(k)\frac{\sum_{j=1}^{m(k)} \left(\frac{\rho k}{C\av{k}}\right)^j/(j-1)!}{\sum_{j=0}^{m(k)}\left(\frac{\rho k}{C\av{k}}\right)^{j}/j!}=\rho,
\end{equation}
an equation which also appeared in \cite{Capacity}. Clearly, for $m(k)\rightarrow \infty$, $C\to 1$, and we reproduce the results of Section \ref{networks_sect}. Eq. \eqref{fj_limited_capacity} (for $j=0$) is compared to simulations in Figure \ref{figure8_freeSites_Capacity_1particle_SFN}.

\begin{figure}[ht]
\centering
\subfigure[]{
   \includegraphics[width=8cm] {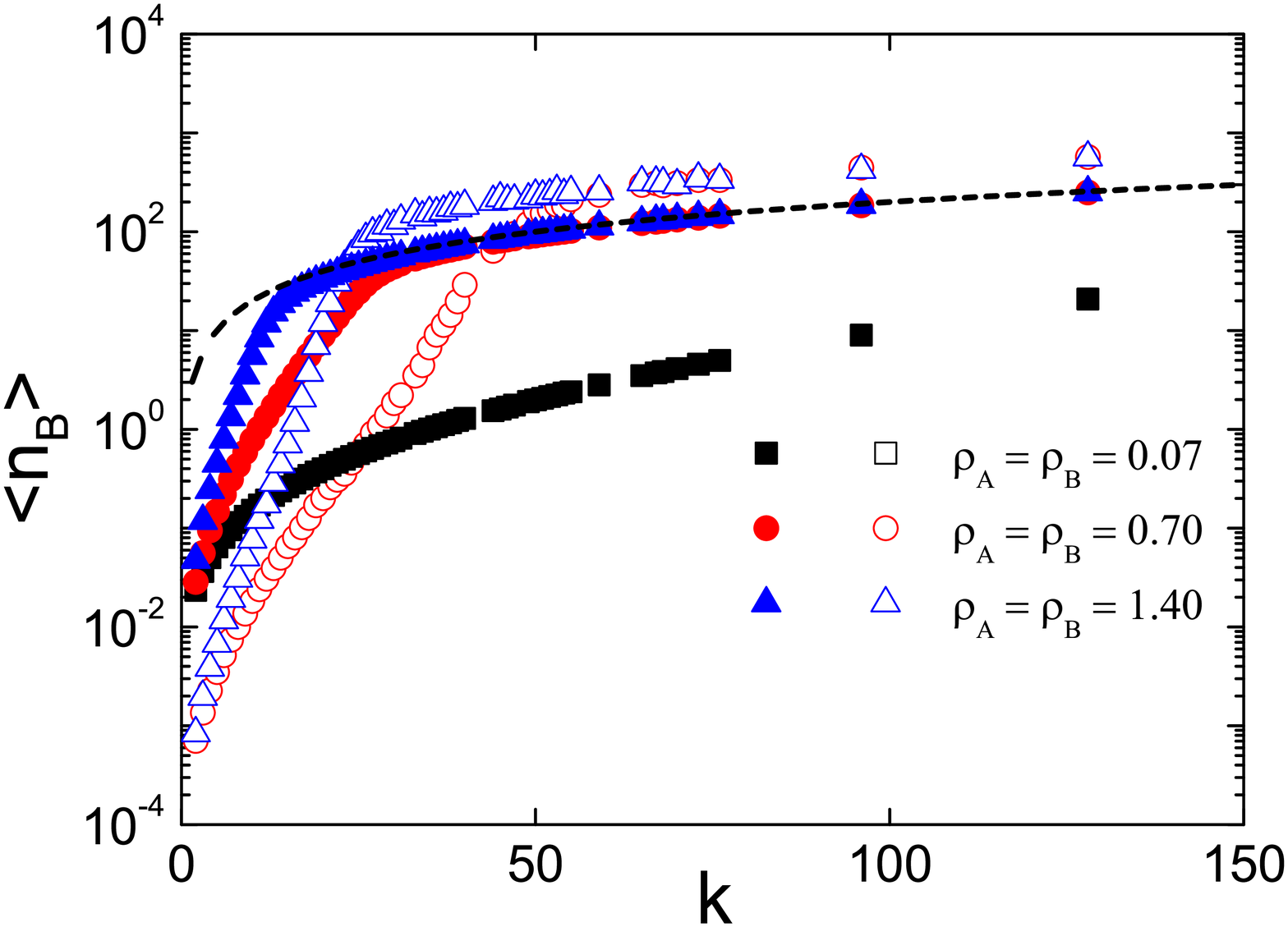}
   \label{figure9_nB_commonBuffer_SFN_capacity2}
}
\subfigure[]{
   \includegraphics[width=8cm] {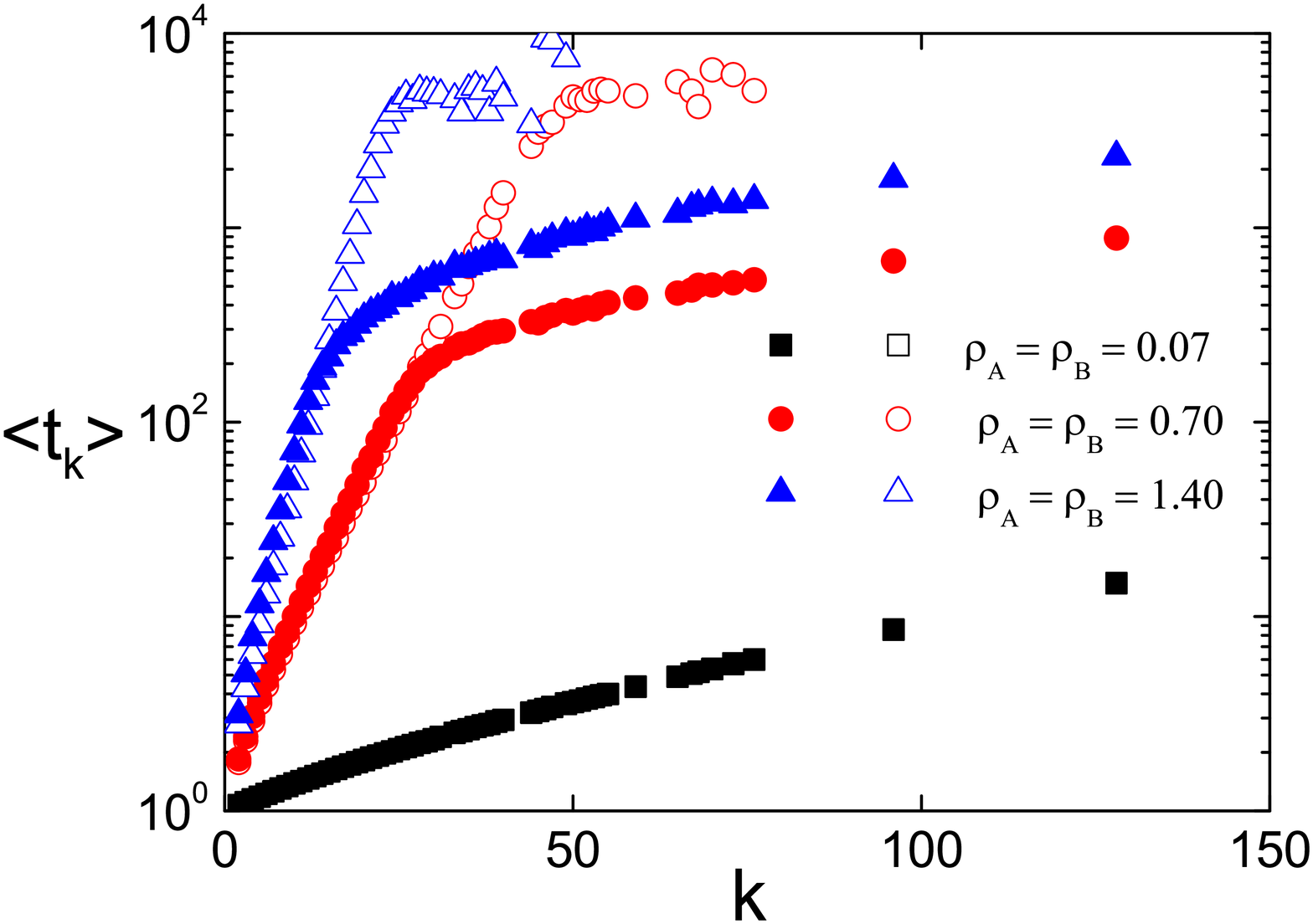}
   \label{figure9_meanTime_commonBuffer_SFN_capacity2}
}
\caption{(Color online) The priority diffusion model with limited node capacity. (a) The average number of $B$ particles, $\av{n_B}$, vs. $k$, for the case of limited node capacity with $m(k)=2k$. The black dashed line corresponds to the capacity, $n_B=m(k)$ (b) The average waiting time ,$\av{t_k}$, vs. $k$. Full symbols are for limited capacity; empty symbols are for infinite capacity.}
\label{figure9_commonBuffer_SFN_capacity2}
\end{figure}

With two species and the priority constraint, a reasonable choice for the capacity is $m(k)\propto k$, since nodes with larger degrees are usually more powerful and can handle more information. As shown in \cite{Capacity} (see also Figure \ref{figure8_freeSites_varCapacity3_1particle_SFN}), the motion of the $A$ particles is not expected to be seriously affected due to the finite capacity. However, we have seen in Section \ref{networks_sect} that in the absence of capacity constraints, the $B$'s concentration grows as $k\exp(\rho_A k/\av{k})$. When finite capacity is imposed, $B$'s
are not able to aggregate at the hubs as before, and their mobility is thus expected to be enhanced. This is demonstrated in Figure \ref{figure9_commonBuffer_SFN_capacity2}, where
we plot the average concentration and the average waiting times for the $B$ particles.

\section{Summary}

In summary, we introduced and analyzed a model of random walk with two species, $A$ and $B$, where
the motion of one species ($A$) has precedence over that of the other. Our analytical results are summarized in Table \ref{SummaryTable}. We obtained expressions for the diffusion coefficients in regular networks and lattices, for three possible
particle selection protocols. In networks, we showed that the key quantity of the number of sites
occupied by low-priority particles only decreases exponentially with the site degree. The
consequence of this finding was an exponentially increasing concentration of the low priority
particles in the hubs, followed by extremely long waiting times between consecutive hops. We used
simulations to confirm this picture.

We then studied several strategies to improve the mobility of the low priority particles while maintaining the priority constraint. In the first strategy, we suggested that a selected $B$ that is unable to move will enforce hopping of a co-existing high-priority $A$. This results in the $A$'s being repelled out of sites with many $B$'s and prevention of the $B$'s trapping. The second strategy was to allow $B$ particles to jump ahead of the $A$'s with a small probability. We obtained the diffusion coefficients of the two species in lattices and showed that in networks, whenever the hopping probability is non-zero, the average waiting time of the $B$'s is finite even at the hubs. We then also considered modifying the nature of the random walk to preferential hopping into non-hub nodes and showed that this strategy distributes the particles more evenly, increasing the chances for a low priority particle to be free to move. Finally, we showed that limiting the queue size at each node can also prohibit the over-crowding of particles at the hubs. We note, however, that in the last two cases, while the waiting times of the low priority particles are shorter, the number of hops they would need to cover the network is expected to increase.
We believe that our analytical and numerical results, for a wide variety of communication protocols and strategies, will be useful for communication network designers whenever protocols involve randomness and priority assignments.

\begin{widetext}

\begin{table}
\begin{tabular}{|c|c|c|c|}
  \hline
  Lattices & $f_0$ & $r$ & $D_B$ \\
  \hline \hline
  Site prot. & $\frac{1}{1+\rho}$ &  & $\frac{\rho_B}{(1+\rho_A)\rho_S}$ \\ \hline
  Redraw prot. & \multirow{2}{*}{$e^{-\rho}$} & $1-\frac{2}{1+\epsilon}\rho_A+\frac{13+2\epsilon-3\epsilon^2}{2(1+\epsilon)^2(2+\epsilon)}\rho_A^2 + {\cal O}(\rho^3)$ & $\frac{\left[r+(1-r)\epsilon\right]\rho_B/\rho_S}{1-(1-r)(1-\epsilon)\rho_B/\rho_S}$ \\ \cline{1-1} \cline{3-4}
  MoveA prot. && $1-\rho_A+\frac{3-\epsilon}{4}\rho_A^2 + {\cal O}(\rho^3)$ & $\left[r + (1-r)\epsilon\right]\rho_B/\rho_S$ \\
  \hline
  \hline
    Networks & Normal & Avoid hubs & Limited capacity \\ \hline
   $f_0$ & $\exp\left(-\rho k/\av{k}\right)$ & $\exp\left(-\rho k^{1+\alpha}/\av{k^{1+\alpha}}\right)$ & $\left[\sum_{j=0}^{m(k)}\left(\frac{\rho k}{C\av{k}}\right)^j/j!\right]^{-1}$ \\ \hline \hline
   $\tau_k$ & $\psi_k(t)$ & $\psi(t)$ & $n_B(k)$ \\ \hline
   $[f_0+\epsilon(1-f_0)]^{-1}$ & $\sim e^{-t/\tau_k}$ & $\sim [\ln^{\gamma-1}t]^{-1}$ & $\sim k\exp\left(\rho k/\av{k}\right)$ \\
  \hline
\end{tabular}
\caption{The analytical results derived in this paper. For lattices, three protocols were considered: a site selection protocol and a particle selection protocol with either \emph{redraw} or \emph{moveA} subprotocols when a $B$ is selected in a site in which $A$s are also present. We calculated the fraction of empty sites, $f_0$, for one species of density $\rho$ as well as the diffusion coefficient of the $B$s, $D_B$, for two species of densities $\rho_A$ and $\rho_B$ ($\rho_S=\rho_A+\rho_B$). For the particle protocols, we calculated a low density approximation for $r$, the probability of a $B$ particle to be free. $\epsilon$ is a ``soft priority'' probability to move a $B$ particle in the presence of $A$s. For networks, the first two rows show the fraction of empty sites, $f_0$, either for normal diffusion, or when hubs are avoided, or when the capacity at the nodes is limited. $k$ is the degree; $\av{k}$ is the average degree; $\alpha$ is the degree-preference exponent (sites are visited with probability proportional to $k^{\alpha}$); $m(k)$ is the capacity of a node of degree $k$; and $C$ is a normalization coefficient calculated from Eq. \eqref{C_limited_capacity}. The final two rows provide additional quantities for networks: $\tau_k$ is the average time spent in a node of degree $k$ (here, soft priorities are also included); $\psi_k(t)$ is the distribution of waiting times of $B$ particles at sites of degree $k$; $\psi(t)$ is the distribution of all waiting times; and $n_B(k)$ is the average number of $B$ particles at a node of degree $k$.} \label{SummaryTable}
\end{table}

\end{widetext}

\section*{Acknowledgements}
We acknowledge financial support from the Israel Science Foundation, the DFG, and the EU projects LINC and MULTIPLEX. S. C. acknowledges financial support from the Human Frontier Science Program. N. B. acknowledges financial support from Public Benefit Foundation Alexander S. Onassis.

\section*{Appendix: Shortest path routing}

\begin{figure}[ht]
\centering \includegraphics[width=8cm] {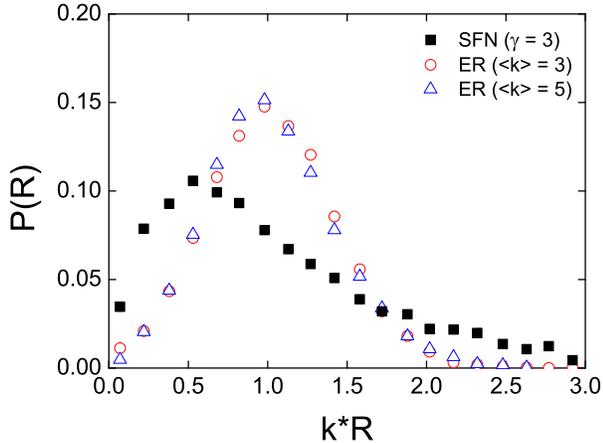}
\caption{(Color online) $P(R)$, The distribution of $R$, the fraction of messages that are routed through a link emerging from a node of degree $k$. Results are averages over 5 realizations for the case of $k=5$. Black squares are for scale-free networks ($\gamma=3,k \geq 2$), open red circles are for ER with $\left\langle k\right\rangle = 3$, and open blue triangles are for ER with $\left\langle k\right\rangle = 5$. Note that the x-axis is scaled by $1/k$. The distribution is narrowly centered around $1/k$ for ER networks, but not for SF networks.}
\label{Appendix_SFN_ER}
\end{figure}

In this appendix, we show that the statistical properties of traffic in homogeneous networks with
shortest path routing resemble, for some networks, those of a random walk. Consider an all-pairs communication model,
where packets are sent from all nodes to all other nodes along shortest paths. Denote the source
node as $i$ and the destination as $j$. At each intermediate node $m$ along the path, the packet
must be sent to the neighbor of $m$ closest to $j$. If the network is homogeneous, we expect the
next node on the path to be, with roughly equal probability, any of the neighbors of $m$, similar to a random walk. To test this, we numerically calculated the fraction of messages routed through each link (which is also the betweeness centrality \cite{centrality}) in our model networks. We compared this quantity, which we call $R$, to $1/k$ ($k$ is the degree of the node from where the message was sent), the probability to route through the link in the case of a random walk. We found that indeed, for the relatively homogeneous regular and ER networks, the probability of routing through a link is narrowly distributed around $1/k$ (Figure \ref{Appendix_SFN_ER}). For the heterogeneous SF networks, the distribution of routing probabilities is wider, since most shortest paths visit specifically the hubs. Thus, as long as the network is homogeneous, our model is expected to describe, at least qualitatively, also the traffic resulting from shortest path routing with priorities.

\bibliographystyle{unsrt}
\bibliography{priority_diffusion}

\end{document}